\documentstyle[pra,aps]{revtex}
\begin{document}

\draft

\title{Theory of spin-2 Bose-Einstein condensates:
spin correlations, magnetic response, and excitation spectra}

\author{Masahito Ueda and Masato Koashi$^1$}

\address{Department of Physics,
Tokyo Institute of Technology, 2-12-1 Ookayama, Meguro-ku,
Tokyo 152-8551, Japan\\
$^1$ CREST Research Team for Interacting Carrier Electronics, School of
Advanced Sciences, The Graduate University for Advanced Studies (SOKEN),
Hayama, Kanagawa 240-0193, Japan
}
\date{\today}

\maketitle

\begin{abstract}
The ground states of Bose-Einstein condensates of spin-2 bosons are     
classified into three distinct (ferromagnetic, ^^ ^^ antiferromagnetic",
and cyclic) phases depending on the s-wave scattering lengths of binary 
collisions for total-spin 0, 2, and 4 channels.
Many-body spin correlations and magnetic response of the condensate in  
each of these phases are studied in a mesoscopic regime, while low-lying
excitation spectra are investigated in the thermodynamic regime.
In the mesoscopic regime, where the system is so tightly confined that  
the spatial degrees of freedom are frozen, the exact, many-body ground  
state for each phase is found to be expressed in terms of the creation  
operators of pair or trio bosons having spin correlations.
These pairwise and trio-wise units are shown to bring about some unique 
features of spin-2 BECs such as a huge jump in magnetization from       
minimum to maximum possible values and the robustness of the 
minimum-magnetization state against an applied magnetic field.
In the thermodynamic regime, where the system is spatially uniform,     
low-lying excitation spectra in the presence of magnetic field are      
obtained analytically using the Bogoliubov approximation.
In the ferromagnetic phase, the excitation spectrum consists of one     
Goldstone mode and four single-particle modes.
In the antiferromagnetic phase, where spin-singlet ^^ ^^ pairs" undergo 
Bose-Einstein condensation, the spectrum consists of two Goldstone modes
and three massive ones, all of which become massless when magnetic field
vanishes.
In the cyclic phase, where boson ^^ ^^ trios" condense into a           
spin-singlet state, the spectrum is characterized by two Goldstone modes,
one single-particle mode having a magnetic-field-independent energy gap,
and a gapless single-particle mode that becomes massless in the absence 
of magnetic field. 
\end{abstract}


\section{Introduction}

Bose-Einstein condensates (BECs) of trapped alkali atoms have internal  
degrees of freedom due to the hyperfine spin of the atoms. 
When a BEC is trapped in a magnetic potential, these degrees of freedom 
are frozen and the state of the BEC is described at a mean-field level  
by a scalar order parameter.
When a BEC is trapped in an optical potential, however, the spin degrees
of freedom are liberated, giving rise to a rich variety of phenomena    
such as spin domains~\cite{Miesner} and textures~\cite{OM}.
Here the order parameter has $2f+1$ components that transform under     
spatial rotation as the spherical tensor of rank $f$, where $\hbar f$ is
the hyperfine spin of bosons.

Mean field theories (MFTs) of spinor BECs were put forth for both      
spin-1~\cite{OM,Ho,Stenger} and spin-2~\cite{Koashi,Ciobanu} cases.    
According to them, the $m=0$ magnetic sublevel of an antiferromagnetic 
BEC is not populated in the presence of magnetic field for both spin-1 
and spin-2 cases.
However, Law et al.~\cite{Law} used many-body theory to show that the  
$m=0$ sublevel of a spin-1 BEC {\it is} macroscopically populated due  
to the formation of spin-singlet ^^ ^^ pairs" of bosons.
It was subsequently shown ~\cite{Koashi} that the $m=0$ sublevel of a  
spin-2 BEC is also macroscopically occupied due to the formation of    
spin-singlet ^^ ^^ trios" of bosons.
The physics common to both cases is that the spin-singlet state is     
isotropic and therefore each magnetic sublevel shares the equal        
population.

A question then arises as to what extent and under what conditions MFTs
are applicable.
It is now understood~\cite{Koashi,Ho2} that the validity of MFTs is    
quickly restored with increasing an applied magnetic field.
Thus for the many-body spin correlations to manifest themselves, the   
external magnetic field have to be very low.

The spin-singlet pairs of bosons should be distinguished from Cooper   
pairs of electrons or those of $^3$He due to the statistical difference
of constituent particles.
The Cooper pairs consist of fermions, so that the state is symmetric   
only under the permutations that do not break any pair.
For the case of spin-singlet pairs of bosons, the state is symmetric   
under any permutation of the constituents.     
The Bose-Einstein statistics leads to a constructive interference among
permuted terms, giving rise to a highly nonlinear magnetic response to 
be discussed later.
In contrast with usual antiferromagnets, where antiparallel spins are  
alternately aligned (Neel order), ^^ ^^ antiferromagnetic" BECs do not 
possess such a long-range spatial order because the system lacks       
crystal order. 
The antiferromagnetic phase of BECs is also called polar~\cite{Ho}.

In Refs.~\cite{Law,Koashi,Ho2}, only spin degrees of freedom are       
considered by assuming that the spatial degrees of freedom are frozen. 
In this paper, we relax this restriction and develop a theory of spin-2
BECs that enables us to study many-body ground states and the excitation
spectrum thereof on an equal footing.
For spin-1 BECs, this program has been carried out in Ref.~\cite{Ueda}.
Many-body spin correlations and magnetic response of BECs, the results 
of which were briefly reported in Ref.~\cite{Koashi}, are expounded.   
The role of symmetry of the ground state in determining the character  
of the excitation spectrum is also elucidated.

This paper is organized as follows.
Section~\ref{sec:Formulation} derives an effective Hamiltonian that    
enables us to study many-body spin correlations and low-lying excitation
spectrum of spin-2 BECs on an equal footing.
Section~\ref{sec:MFT} reviews mean-field properties of each phase of   
BECs.
Section~\ref{sec:MBT} studies many-body spin correlations and magnetic 
response of BEC. 
The energy eigenstate is explicitly constructed using the creation     
operators of boson pairs and trios. 
The degeneracy of the eigenstate is examined and some novel magnetic   
response such as a huge jump in magnetization and the robustness of the
minimum-magnetization state against an applied magnetic field are discussed.
Section~\ref{sec:Bogoliubov} derives excitation spectra of BECs using  
the Bogoliubov approximation.
All excitation spectra are obtained analytically and the relations of  
their characters to the symmetry of the ground state are discussed.    
Section~\ref{sec:Conclusions} summarizes the main results of the present
paper.
Appendix~\ref{app:Characterization} recapitulates the parametrization  
of the order parameter of spin-2 BECs, and appendix~\ref{app:zeeman}   
describes a method of calculating Zeeman-level populations.

\section{Formulation of the problem}
\label{sec:Formulation}

\subsection{Interaction Hamiltonian}

Consider a system of identical bosons with hyperfine spin $f$ and let
$\hat{\Psi}_m(\bbox{r})$ ($m=f,f-1,\cdots,-f$) be the field operator
that annihilates at position $\bbox{r}$ a boson with magnetic
quantum number $m$. The field operators are 
assumed to obey the canonical commutation relations
\begin{eqnarray}
[\hat{\Psi}_m(\bbox{r}),\hat{\Psi}^\dagger_n(\bbox{r}')]
=\delta_{mn}\delta(\bbox{r}-\bbox{r}'),  \ \
[\hat{\Psi}_m(\bbox{r}),\hat{\Psi}_n(\bbox{r}')]=0, \ \
[\hat{\Psi}_m^\dagger(\bbox{r}),\hat{\Psi}_n^\dagger(\bbox{r}')]
=0,
\label{com}
\end{eqnarray}
where the Kronecker's delta $\delta_{mn}$ takes on the value of 1 
if $m=n$ and 0 otherwise.
The Bose-Einstein statistics requires that the total spin of any two 
bosons whose relative orbital angular momentum is zero be restricted to
$F=2f,2f-2,\cdots,0$. We may therefore use $F$ as an index for
classifying binary interactions between identical bosons:
\begin{eqnarray}
\hat{V}=\sum_{F=0,2,\cdots,2f}\hat{V}^{(F)},
\label{int1}
\end{eqnarray}
where $\hat{V}^{(F)}$ describes an interaction between two bosons
whose total spin is $F$. To construct $\hat{V}^{(F)}$, consider the
operator $\hat{A}_{FM}(\bbox{r},\bbox{r}')$ that annihilates at 
positions $\bbox{r}$ and $\bbox{r}'$ two bosons
with total spin $F$ and total magnetic quantum number $M$:
\begin{eqnarray}
\hat{A}_{FM}(\bbox{r},\bbox{r}')=\sum_{m_1,m_2=-f}^f
\langle F,M|f,m_1;f,m_2\rangle
\hat{\Psi}_{m_1}(\bbox{r})\hat{\Psi}_{m_2}(\bbox{r}'),
\label{A}
\end{eqnarray}
where $\langle F,M|f,m_1;f,m_2\rangle$ is the Clebsch-Gordan coefficient.
We may use $\hat{A}_{FM}$ to construct $\hat{V}^{(F)}$ as
\begin{eqnarray}
\hat{V}^{(F)}=\frac{1}{2}
\int \!d\bbox{r}\int \!d\bbox{r}' v^{(F)}(\bbox{r},\bbox{r}')\sum_{M=-F}^{F}
\hat{A}_{FM}^\dagger(\bbox{r},\bbox{r}')\hat{A}_{FM}(\bbox{r},\bbox{r}'),
\label{int2}
\end{eqnarray}
where $v^{(F)}(\bbox{r},\bbox{r}')$ describes the dependence of the
interaction on the positions of the particles.
Because of the completeness relation
$\sum_{F,M}
|F,M\rangle\langle F,M|=\hat{1}$, where $\hat{1}$ is the identity
operator, we find that
\begin{eqnarray}
\sum_{F=0,2,\cdots,2f}\sum_{M=-F}^F
\hat{A}_{FM}^\dagger(\bbox{r},\bbox{r}')\hat{A}_{FM}(\bbox{r},\bbox{r}')
&=&
:\hat{n}(\bbox{r})\hat{n}(\bbox{r}'):,
\label{sum1}
\end{eqnarray}
where
\begin{eqnarray}
\hat{n}(\bbox{r})\equiv\sum_{m=-f}^f\hat{\Psi}_m^\dagger(\bbox{r})
\hat{\Psi}_m(\bbox{r})
\end{eqnarray}
is the total density operator and
:: denotes normal ordering, that is, annihilation operators are
placed to the right of creation operators.
Integrating Eq.~(\ref{sum1}) over $\bbox{r},\bbox{r}'$ yields
\begin{eqnarray}
\int \!d\bbox{r}\int \!d\bbox{r}'
\sum_{F,M}\hat{A}_{FM}^\dagger(\bbox{r},\bbox{r}')\hat{A}_{FM}(\bbox{r},{\bf
r}')
=\hat{N}(\hat{N}-1),
\label{sum2}
\end{eqnarray}
where $\hat{N}\equiv \int \!\hat{n}(\bbox{r})d\bbox{r}$ is the total number
of bosons.

In the case of a dilute Bose-Einstein condensate of neutral atoms,
we may to a good approximation assume that
$v^{(F)}(\bbox{r},\bbox{r}')=g_F\delta(\bbox{r}-\bbox{r}')$,
where $g_F$ characterizes the strength of the interaction between two bosons
whose total spin is $F$, and is related to the corresponding s-wave
scattering length $a_F$ as 
\begin{eqnarray}
g_F=\frac{4\pi\hbar^2}{M}a_F.
\label{g_F}
\end{eqnarray}
Equation~(\ref{int2}) then becomes
\begin{eqnarray}
\hat{V}^{(F)}=\frac{g_F}{2}\int \!d\bbox{r}\sum_{M=-F}^{F}
\hat{A}_{FM}^\dagger(\bbox{r},\bbox{r})\hat{A}_{FM}(\bbox{r},\bbox{r}).
\label{int4}
\end{eqnarray}
In the following discussions we shall focus on this case and 
therefore denote $\hat{A}_{FM}(\bbox{r},\bbox{r})$ simply as
$\hat{A}_{FM}(\bbox{r})$.

When $f=2$, $F$ can take on values 0, 2, and 4.
For $F=0$, we have
\begin{eqnarray}
\hat{V}^{(0)}=\frac{g_0}{2}
\int \!d\bbox{r}
\hat{A}_{00}^\dagger(\bbox{r})\hat{A}_{00}(\bbox{r}),
\label{spin2F=0}
\end{eqnarray}
where
\begin{eqnarray}
\hat{A}_{00}(\bbox{r})
&=&\frac{1}{\sqrt{5}}\left[
2\hat{\Psi}_{2}(\bbox{r})\hat{\Psi}_{-2}(\bbox{r})
-2\hat{\Psi}_{1}(\bbox{r})\hat{\Psi}_{-1}(\bbox{r})
+\hat{\Psi}_0^2(\bbox{r})
\right].
\label{spin2A00}
\end{eqnarray}
For $F=2$, we have
\begin{eqnarray}
\hat{V}^{(2)}=\frac{g_2}{2}\int \!d\bbox{r}\sum_{M=-2}^{2}
\hat{A}_{2M}^\dagger(\bbox{r})\hat{A}_{2M}(\bbox{r}).
\label{spin2F=2}
\end{eqnarray}
For $F=4$ we have
\begin{eqnarray}
\hat{V}^{(4)}&=&\frac{g_4}{2}\int \!d\bbox{r}
\sum_{M=-4}^{4}
\hat{A}_{4M}^\dagger(\bbox{r})\hat{A}_{4M}(\bbox{r})
\nonumber \\
&=&\frac{g_4}{2}\int \!d\bbox{r}\left[
:\hat{n}^2(\bbox{r}):
-\hat{A}_{00}^\dagger(\bbox{r})\hat{A}_{00}(\bbox{r})
-\sum_{M=-2}^2
\hat{A}_{2M}^\dagger(\bbox{r})\hat{A}_{2M}(\bbox{r})
\right],
\label{spin2F=4}
\end{eqnarray}
where Eq.~(\ref{sum1}) was used in obtaining the second equality.
Summing Eqs.~(\ref{spin2F=0}), (\ref{spin2F=2}) and (\ref{spin2F=4}), we
obtain the interaction Hamiltonian as
\begin{eqnarray}
\hat{V}&=&\hat{V}^{(0)}+\hat{V}^{(2)}+\hat{V}^{(4)}
\nonumber \\
&=&\frac{1}{2}\int \!d\bbox{r}
\left[
g_4:\hat{n}^2(\bbox{r}):
+(g_0-g_4)
\hat{A}_{00}^\dagger(\bbox{r})\hat{A}_{00}(\bbox{r})
+(g_2-g_4)\sum_{M=-2}^2
\hat{A}_{2M}^\dagger(\bbox{r})\hat{A}_{2M}(\bbox{r})
\right].
\label{int}
\end{eqnarray}
To eliminate the last term in Eq.~(\ref{int}), we note the following
operator identity:
\begin{eqnarray}
\frac{1}{7}:\hat{\bbox{F}}^2(\bbox{r}):+
\sum_{M=-2}^2
\hat{A}_{2M}^\dagger(\bbox{r})\hat{A}_{2M}(\bbox{r})
+\frac{10}{7}\hat{A}_{00}^\dagger(\bbox{r})\hat{A}_{00}(\bbox{r})
=\frac{4}{7}:\hat{n}^2(\bbox{r}):,
\label{spin2identity}
\end{eqnarray}
where $\hat{\bbox{F}}=(\hat{F}^x,\hat{F}^y,\hat{F}^z)$ represents the spin
density operators defined by
\begin{eqnarray}
\hat{F}_i(\bbox{r})=\sum_{m,n=-1}^1 f^i_{mn}\hat{\Psi}_m^\dagger(\bbox{r})
\hat{\Psi}_n(\bbox{r}) \ (i=x,y,z)
\label{spindensity}
\end{eqnarray}
with $f^i_{mn}$ $(i=x,y,z)$ being the $(m,n)$-components of
spin-2 matrices $f^i$ given by
\begin{eqnarray}
f^x=\frac{1}{2}
\left(
\begin{array}{ccccc}
0 & 2        & 0        & 0        & 0 \\
2 & 0        & \sqrt{6} & 0        & 0 \\
0 & \sqrt{6} & 0        & \sqrt{6} & 0 \\
0 & 0        & \sqrt{6} & 0        & 2 \\
0 & 0        & 0        & 2        & 0
\end{array}
\right), \ \ \
f^y=\frac{i}{2}
\left(
\begin{array}{ccccc}
0 & -2       & 0        & 0        & 0 \\
2 & 0        & -\sqrt{6}& 0        & 0 \\
0 & \sqrt{6} & 0        & -\sqrt{6}& 0 \\
0 & 0        & \sqrt{6} & 0        & -2\\
0 & 0        & 0        & 2        & 0
\end{array}
\right), \ \ \
f^z=
\left(
\begin{array}{ccccc}
2 & 0 & 0 & 0 & 0 \\
0 & 1 & 0 & 0 & 0 \\
0 & 0 & 0 & 0 & 0 \\
0 & 0 & 0 & -1& 0 \\
0 & 0 & 0 & 0 & -2
\end{array}
\right).
\label{spin-2_mat}
\end{eqnarray}
We may use Eq.~(\ref{spin2identity}) to eliminate the last term in
Eq.~(\ref{int}), obtaining
\begin{eqnarray}
\hat{V}
&=&\frac{1}{2}\int \!d\bbox{r}
\left[
      c_0:\hat{n}^2(\bbox{r}):
+ c_1:\hat{\bbox{F}}^2(\bbox{r}):
+ c_2\hat{A}_{00}^\dagger(\bbox{r})
\hat{A}_{00}(\bbox{r})
\right],
\label{spin2int2}
\end{eqnarray}
where $c_0\equiv(4g_2+3g_4)/7$, $c_1\equiv(g_4-g_2)/7$, and
$c_2\equiv(7g_0-10g_2+3g_4)/7$.

\subsection{Total Hamiltonian}

In the following discussions we shall assume that the external magnetic
field is weak enough to ignore the quadratic Zeeman effect.
Then the total Hamiltonian $\hat{H}$ of the system consists of
the kinetic energy term $\hat{H}_{\rm KE}$,
the trapping potential energy term $\hat{H}_{\rm PE}$,
the linear Zeeman term $\hat{H}_{\rm LZ}$,
and the interaction term $\hat{V}$:
\begin{eqnarray}
\hat{H}_{\rm KE}&=&\int d\bbox{r} \sum_{m=-2}^2
\hat{\Psi}_m^\dagger
\left(-\frac{\hbar^2\nabla^2}{2M}\right)\hat{\Psi}_m,
\label{KE} \\
\hat{H}_{\rm PE}&=&\int d\bbox{r} \sum_{m=-2}^2
\hat{\Psi}_m^\dagger U_{\rm trap}(\bbox{r})\hat{\Psi}_m,
\label{PE} \\
\hat{H}_{\rm LZ}
&=&-p\int d\bbox{r}\sum_{m,n=-2}^2
\hat{\Psi}_m f^z_{mn}\hat{\Psi}_n
=-p\int d\bbox{r}\sum_{m=-2}^2
m \hat{\Psi}_m^\dagger\hat{\Psi}_m,
\label{LZ}
\end{eqnarray}
where $p$ $(>0)$ is the product of the gyromagnetic ratio
and the external magnetic field which is assumed to be applied 
in the z-direction.

When the system is spatially uniform, i.e. $U_{\rm trap}(\bbox{r})=0$,
it is convenient to expand the field operator in terms of
plane waves:
\begin{eqnarray}
\hat{\Psi}_m(\bbox{r})=\frac{1}{\sqrt{V}}\sum_{\bbox{k}}\hat{a}_{{\bbox{k}}m
}
e^{i{\bbox{k}\cdot\bbox{r}}}.
\label{PWE}
\end{eqnarray}
Then the single-particle part of the Hamiltonian becomes
\begin{eqnarray}
\hat{H}_0&=&\hat{H}_{\rm KE}+\hat{H}_{\rm LZ}
=\sum_{{\bbox{k}},m}
(\epsilon_{\bbox{k}}-pm)\hat{n}_{{\bbox{k}},m},
\label{H0}
\end{eqnarray}
where $\epsilon_{\bbox{k}}\equiv\hbar^2k^2/2M$,
and the interaction Hamiltonian becomes
\begin{eqnarray}
\hat{V}=\frac{1}{2V}
\sum_{\bbox{k}}:(c_0\hat{\rho}_{\bbox{k}}^\dagger\hat{\rho}_{\bbox{k}}
+c_1\hat{\bbox{F}}_{\bbox{k}}^\dagger\cdot\hat{\bbox{F}}_{\bbox{k}}
+c_2\hat{A}_{\bbox{k}}^\dagger\hat{A}_{\bbox{k}}
):,
\label{spin-1V1}
\end{eqnarray}
where
\begin{eqnarray}
& & \hat{\rho}_{\bbox{k}}
\equiv\int  \hat{n}(\bbox{r})e^{-i{\bbox{k}\cdot\bbox{r}}} d\bbox{r}
=\sum_{\bbox{q}m}\hat{a}_{\bbox{q},m}^\dagger\hat{a}_{\bbox{q}+{\bbox{k}},m},
\\
& & \hat{\bbox{F}}_{\bbox{k}}
\equiv \int  \hat{\bbox{F}}(\bbox{r})e^{-i{\bbox{k}\cdot\bbox{r}}}
d\bbox{r} =\sum_{\bbox{q}mn}{\bf f}_{mn}
\hat{a}_{\bbox{q},m}^\dagger\hat{a}_{\bbox{q}+{\bbox{k}},n}, \\
& & \hat{A}_{\bbox{k}}
\equiv \int \hat{A}_{00}(\bbox{r})e^{-i{\bbox{k}\cdot\bbox{r}}} d\bbox{r}
=\frac{1}{\sqrt{5}}\sum_{\bbox{q},m}(-1)^m\hat{a}_{\bbox{q},m}
\hat{a}_{{\bbox{k}-\bbox{q}},-m}.
\end{eqnarray}

\section{Mean-field theory}
\label{sec:MFT}

Mean field theory (MFT) of spin-2 BEC was discussed in
Refs.~\cite{Koashi,Ciobanu}.
We present here a brief summary of as much of this theory as is relevant 
to later discussions. We shall also present some new results.
When the system is uniform, BEC occurs in the ${\bbox{k}}={\bf 0}$ state.
We therefore assume the following trial wave function
\begin{eqnarray}
|{\bbox \zeta}\rangle=
\frac{1}{\sqrt{N!}}\left(\sum_{m=-2}^2\zeta_m
\hat{a}_{{\bf 0},m}^\dagger
\right)^{N}\!
|{\rm vac}\rangle,
\label{SPS}
\end{eqnarray}
where the complex amplitudes $\zeta_m$ are assumed to satisfy 
the normalization condition
\begin{eqnarray}
\sum_{m=-2}^2|\zeta_m|^2=1.
\label{norm}
\end{eqnarray}
The variational parameters $\zeta_m$ are determined 
so as to minimize the expectation value of
$\hat{H}$ over the state (\ref{SPS}):
\begin{eqnarray}
\langle\hat{H}\rangle=\frac{c_0}{2V}N(N-1)+
\frac{c_1}{2V}N(N-1)\langle\hat{\bbox{f}}\rangle^2
+\frac{2c_2}{5V}N(N-1)|\langle\hat{s}_-\rangle|^2
-pN\langle\hat{f}_z\rangle,
\label{Ham}
\end{eqnarray}
where
\begin{eqnarray}
& & 
\langle\hat{\bbox{f}}\rangle\equiv\sum_{mn}{\bbox{f}}_{mn}\zeta_m^*\zeta_n,
\label{f} \\
& & \langle\hat{f}_z\rangle\equiv\sum_{m}m|\zeta_m|^2,
\label{f_z} \\
& & \langle\hat{s}_-\rangle\equiv\frac{1}{2}\sum_m(-1)^m\zeta_m\zeta_{-m} .
\label{s-}
\end{eqnarray}
When the external magnetic field is applied in the $z$-direction, only the
$z$-component of $\langle \hat{\bbox{f}}\rangle$ is nonzero.
We thus obtain
\begin{eqnarray}
\langle\hat{H}\rangle
\equiv\frac{c_0}{2V}N(N-1)+\frac{N(N-1)}{2V}\epsilon_0,
\end{eqnarray}
where
\begin{eqnarray}
\epsilon_0=
c_1\langle\hat{f}_z\rangle^2+\frac{4}{5}c_2|\langle\hat{s}_-\rangle|^2
-\tilde{p}\langle\hat{f}_z\rangle
\label{e}
\end{eqnarray}
with $\tilde{p}\equiv 2Vp/(N-1)$.

The mean-field solution should be determined so as to minimize
$\epsilon_0$ subject to the normalization condition (\ref{norm}):
\begin{eqnarray}
\frac{\partial}{\partial\zeta_m^*}
\left(\epsilon_0-\lambda\sum_m|\zeta_m|^2\right)
=(2c_1\langle\hat{f}_z\rangle-\tilde{p})m\zeta_m
-\lambda\zeta_m+\frac{4}{5}c_2(-1)^m
\langle\hat{s}_-\rangle\zeta_{-m}^*=0,
\label{cond1}
\end{eqnarray}
where $\lambda$ is a Lagrange multiplier.
Multiplying both sides by $(-1)^m\zeta_{-m}$ and summing over $m$ yield
\begin{eqnarray}
\left(\lambda-\frac{2}{5}c_2\right)\langle\hat{s}_-\rangle=0.
\label{cond2}
\end{eqnarray}
On the other hand, multiplying both sides of Eq.~(\ref{cond1})
by $\zeta_m^*$ and summing over $m$ yields
\begin{eqnarray}
(2c_1\langle\hat{f}_z\rangle-\tilde{p})\langle\hat{f}_z\rangle
-\lambda+\frac{8}{5}c_2|\langle\hat{s}_-\rangle|^2=0.
\label{cond3}
\end{eqnarray}

\subsection{Ferromangetic BEC}
\label{sec:MFFM}

When $\langle\hat{s}_-\rangle=0$, Eq.~(\ref{cond1}) gives
\begin{eqnarray}
\left[(2c_1\langle\hat{f}_z\rangle-\tilde{p})m-\lambda
\right]\zeta_m=0.
\label{fcond}
\end{eqnarray}
For nonzero components $\zeta_m\neq0$, Eq.~(\ref{fcond}) gives
\begin{eqnarray}
(2c_1\langle\hat{f}_z\rangle-\tilde{p})m-\lambda=0.
\label{fcond2}
\end{eqnarray}
The case of $\tilde{p}=2c_1\langle\hat{f}_z\rangle$ will be discussed
in Sec.~\ref{sec:CBEC}.
When $\tilde{p}\neq2c_1\langle\hat{f}_z\rangle$, only one component can
be nonzero.
The mean-field solutions, magnetizations, and mean-field energies are
therefore given by
\begin{eqnarray}
& & {\bbox \zeta}=e^{i\phi}(1,0,0,0,0), \ \ \langle\hat{f}_z\rangle=2, \ \
\epsilon^{\rm F}=4c_1-2\tilde{p}, \label{OP_F} \\
& & {\bbox \zeta}=e^{i\phi}(0,1,0,0,0), \ \ \langle\hat{f}_z\rangle=1, \ \
\epsilon^{\rm F}=c_1-\tilde{p}, \label{OP_F2} \\
& & {\bbox \zeta}=e^{i\phi}(0,0,0,1,0), \ \ \langle\hat{f}_z\rangle=-1, \
\
\epsilon^{\rm F}=c_1+\tilde{p}, \label{OP_F3} \\
& & {\bbox \zeta}=e^{i\phi}(0,0,0,0,1), \ \ \langle\hat{f}_z\rangle=-2, \
\
\epsilon^{\rm F}=4c_1+2\tilde{p}, \label{OP_F4}
\end{eqnarray}
where $\phi$ is an arbitrary global phase.

The ground state is degenerate with respect to the global 
phase $\phi$. This represents the gauge invariance, i.e. conservation
of the total number of particle and leads to a massless Goldstone mode,
as will be shown in Sec.~\ref{sec:FMB}. The conservation of the spin
angular momentum does not lead to a new Goldstone mode because in
ferromagnets all spins are aligned in the same direction and therefore
the total spin angular momentum has the same piece of information as
the total number of particles.

\subsection{Antiferromagnetic BEC}
\label{sec:MFAF}

The antiferromagnetic (or polar) phase of a BEC is defined as the one 
having nonzero spin-singlet pair amplitude, $\langle\hat{s}_-\rangle\neq0$.
When $\langle\hat{s}_-\rangle\neq0$, Eq.~(\ref{cond2}) gives
$\lambda=2c_2/5$. Substituting this and Eq.~(\ref{cond3}) into
Eq.~(\ref{e}) yields
\begin{eqnarray}
\epsilon_0^{\rm
AF}=\frac{c_2}{5}-\frac{\tilde{p}}{2}\langle\hat{f}_z\rangle.
\label{e2}
\end{eqnarray}
With $\lambda=2c_2/5$, Eq.~(\ref{cond1}) leads to
\begin{eqnarray}
(2c_1\langle\hat{f}_z\rangle-\tilde{p})
\left[\left(2c_1m^2-\frac{2}{5}c_2\right)\langle\hat{f}_z\rangle
-\tilde{p}m^2\right]\zeta_m=0.
\label{cond4}
\end{eqnarray}

When $\tilde{p}\neq2c_1\langle\hat{f}_z\rangle$, the solutions of
Eq.~(\ref{cond4}) is that only ($\zeta_2$, $\zeta_{-2}$) or
($\zeta_1$, $\zeta_{-1}$) or $\zeta_0$ is nonzero.
Determining these values using conditions
$\langle\hat{f}_z\rangle=\sum_mm|\zeta_m|^2$
and Eq.~(\ref{norm}), we obtain the mean-field solutions and the
corresponding magnetizations as~\cite{Ciobanu}
\begin{eqnarray}
{\bbox \zeta}&=\frac{1}{\sqrt{2}}e^{i\phi}
\left(\sqrt{1+\frac{\langle\hat{f}_z\rangle}{2}}, 0, 0, 0,
e^{i\chi}\sqrt{1-\frac{\langle\hat{f}_z\rangle}{2}}\right), \ \
& \langle\hat{f}_z\rangle=\frac{2\tilde{p}}{4c_1-c_2/5}, \ \ \ \
\epsilon^{\rm AF}=\frac{c_2}{5}-\frac{\tilde{p}^2}{4c_1-c_2/5}
\label{AFspinor} \\
{\bbox \zeta}&=\frac{1}{\sqrt{2}}e^{i\phi}
\left(0,\sqrt{1+\langle\hat{f}_z\rangle}, 0,
e^{i\chi_1}\sqrt{1-\langle\hat{f}_z\rangle},0\right), \ \
& \langle\hat{f}_z\rangle=\frac{\tilde{p}}{2(c_1-c_2/5)}, \ \
\epsilon^{\rm AF}=\frac{c_2}{5}-\frac{\tilde{p}^2}{4(c_1-c_2/5)}
\label{AFspin2} \\
{\bbox \zeta}&=e^{i\phi}
\left(0, 0, 1, 0,0
\right), \ \
& \langle\hat{f}_z\rangle=0, \ \ \ \ \ \ \ \ \ \
\epsilon^{\rm AF}=\frac{c_2}{5}.
\label{AFspin3}
\end{eqnarray}

The mean-field solutions (\ref{AFspinor}) and (\ref{AFspin2}) are
degenerate with respect to two continuous phase variables, that is,
the global phase $\phi$ and the relative phase 
$\chi_m=\phi_{-m}-\phi_m$
($m=1,2$) between the two nonvanishing amplitudes $\zeta_{\pm m}$.
Corresponding to these two continuous degeneracies, we expect to
have two Goldstone modes, as will be shown in Sec.~\ref{sec:AFB}.

When $\tilde{p}=2c_1\langle\hat{f}_z\rangle$,
Eq.~(\ref{cond3}) with $\lambda=2c_2/5$ gives
$|\langle\hat{s}_-\rangle|=1/2$,
which, together with (\ref{cond1}), leads to
$\zeta_m=e^{2i\phi_0}(-1)^m\zeta^*_{-m}$. Hence we have
$\langle\hat{f}_z\rangle=0$.
This is possible only when the external magnetic field is zero.
The corresponding order parameter is given by
\begin{eqnarray}
\bbox{\zeta}=e^{i\phi_0}
(\frac{e^{i(\phi_2-\phi_0)}}{\sqrt{2}}\sin\theta\sin\psi ,
    \frac{e^{i(\phi_1-\phi_0)}}{\sqrt{2}}\sin\theta\cos\psi ,
    \cos\theta,
-\frac{e^{-i(\phi_1-\phi_0)}}{\sqrt{2}}\sin\theta\cos\psi ,
    \frac{e^{-i(\phi_2-\phi_0)}}{\sqrt{2}}\sin\theta\sin\psi ).
\label{AFspinor2}
\end{eqnarray}
This solution is degenerate with respect to five continuous variables:
one global gauge ($\phi_0$), two relative gauges ($\phi_2-\phi_0$ and
$\phi_1-\phi_0$), and two variables $\theta$ and $\psi$ that specify
the amplitudes of the order parameter. 
As a consequence of these degeneracies, we
expect to have five (three density-like and two spin-like) Goldstone
modes, as will be shown in Sec.~\ref{sec:AFB}.

\subsection{Cyclic BEC}
\label{sec:CBEC}

The remaining possibility is the case in which
$\langle\hat{s}_-\rangle=0$ and
$\langle\hat{f}_z\rangle=\tilde{p}/2c_1$.
This phase will be referred to as cyclic phase.
The energy of this phase is given from Eq.~(\ref{e}) by
\begin{eqnarray}
\epsilon_0^{\rm C}=
-c_1\langle\hat{f}_z\rangle^2.
\label{cyclic}
\end{eqnarray}

Let us now parameterize the order parameter of the cyclic phase as it will
be needed to find the Bogoliubov spectrum.
There are four equations (six real equations) that restrict the order 
parameter of this phase, that is,
\begin{eqnarray}
& & \sum_m|\zeta_m|^2=1,
\label{sum}
\\
& & \langle\hat{s}_-\rangle=\frac{1}{2}\sum_m(-1)^m\zeta_m\zeta_{-m}=0,
\label{S-}
\\
& & \langle\hat{f}_z\rangle=\sum_mm|\zeta_m|^2=\frac{\tilde{p}}{2c_1},
\label{fz}
\\
& & \langle\hat{f}_+\rangle=2(\zeta_2^*\zeta_1+\zeta_{-1}^*\zeta_{-2})
+\sqrt{6}(\zeta_1^*\zeta_0+\zeta_0^*\zeta_{-1})=0.
\label{f+}
\end{eqnarray}
We use the representation (\ref{order}) of the order parameter derived
in Appendix~\ref{app:Characterization} to analyze the cyclic phase. 
This representation automatically satisfies the normalization condition
(\ref{sum}).
To meet the condition (\ref{S-}), we note that
\begin{eqnarray}
\langle\hat{s}_-\rangle=\frac{1}{2}{\rm Tr}{\bf M}^2
=\frac{1}{2}\zeta_0^2-\zeta_1\zeta_{-1}+\zeta_2\zeta_{-2}
=\frac{i}{2}\sin\chi\cos(\phi-\theta).
\end{eqnarray}
The condition (\ref{S-}) therefore requires either
\begin{eqnarray}
\chi=0 \ {\rm or} \  \phi-\theta=\pi/2.
\label{cycond1}
\end{eqnarray}
On the other hand, Eq.~(\ref{f+}) becomes
\begin{eqnarray}
\langle\hat{f}_+\rangle=2\sin\delta\cos\chi
\left[\cos\psi\sin(\theta+\pi/6)-i\sin\psi\sin(\theta-\pi/6)\right]=0,
\end{eqnarray}
whence we obtain
\begin{eqnarray}
{\rm (i)}   \ \chi=\pi/2, \  {\rm or} \
{\rm (ii)}  \  \delta=0, \ {\rm or} \
{\rm (iii)} \ \psi=\pi/2 \ {\rm and} \ \theta=\pi/6, \ {\rm or} \
{\rm (iv)}  \ \psi=0 \ {\rm  and} \  \theta=-\pi/6.
\label{cycond2}
\end{eqnarray}
   From conditions (\ref{cycond1}) and (\ref{cycond2}), we find the
following three solutions and the corresponding magnetization:
\begin{eqnarray}
\zeta_{\pm2}&=\frac{1}{2}(\cos\theta\pm\cos\chi-i\sin\theta\sin\chi)e^{i\phi}, \
\zeta_{\pm1}=0, \
\zeta_0=\frac{1}{\sqrt{2}}(\sin\theta+i\cos\theta\sin\chi)e^{i\phi}, \
& \langle\hat{f}_z\rangle=2\cos\chi\cos\theta,
\label{mfs1} \\
\zeta_{\pm2}&=\frac{1}{2}\left(\frac{\sqrt{3}}{2}\pm\cos\delta\cos\chi
-\frac{i}{2}\sin\chi\right)e^{i\phi}, \
\zeta_{\pm1}=\pm\frac{i}{2}e^{i\phi}\sin\delta\cos\chi , \
\zeta_0=\frac{1}{\sqrt{2}}\left(\frac{1}{2}+i\frac{\sqrt{3}}{2}\sin\chi
\right)e^{i\phi}, \
& \langle\hat{f}_z\rangle=\sqrt{3}\cos\delta\cos\chi,
\label{mfs2} \\
\zeta_{\pm2}&=\frac{1}{2}\left(\frac{\sqrt{3}}{2}\pm\cos\delta\cos\chi
+\frac{i}{2}\sin\chi\right)e^{i\phi}, \
\zeta_{\pm1}=\frac{1}{2}e^{i\phi}\sin\delta\cos\chi , \
\zeta_0=\frac{1}{\sqrt{2}}\left(-\frac{1}{2}+i\frac{\sqrt{3}}{2}
\sin\chi\right)e^{i\phi}, \
& \langle\hat{f}_z\rangle=\sqrt{3}\cos\delta\cos\chi,
\label{mfs3}
\end{eqnarray}
where a global phase $\phi$, which is chosen to be a particular value 
in Eq.(\ref{Re}), is recovered.
While these solutions include three parameters, 
the condition (\ref{fz}) leaves only two parameters free.
It should be noted that these three solutions give the same 
ground-state energy, and
hence are equally possible unless magnetization exceeds $\sqrt{3}$.
When it exceeds $\sqrt{3}$, only solution (\ref{mfs1}) is possible.

A remark is here in order.
As the representation (\ref{order}) is obtained by assuming the full 
isotropy of space, it does not cover the whole order parameter space 
in the presence of magnetic field. We should therefore keep in mind  
that the solutions (\ref{mfs1})-(\ref{mfs3}) do not necessarily exhaust
the whole order parameter space of the cyclic phase.

In the following discussions we shall focus on the solution (\ref{mfs1}).
Making the absolute square of each amplitude yields
\begin{eqnarray}
|\zeta_{\pm2}|^2=\frac{1}{4}(1\pm\langle\hat{f}_z\rangle/2)^2, \ \
|\zeta_0|^2=\frac{1}{2}(1-\langle\hat{f}_z\rangle^2/4).
\end{eqnarray}
Hence we obtain~\cite{Ciobanu}
\begin{eqnarray}
\zeta_{\pm2}=\frac{1}{2}\left(1\pm\frac{\langle\hat{f}_z\rangle}{2}\right)e^
{i\phi_{\pm2}}, \ \
\zeta_{\pm1}=0, \ \
\zeta_0=\sqrt{\frac{1-\langle\hat{f}_z\rangle^2/4}{2}}e^{i\phi_0}, \ \
\label{cyclic_mfs}
\end{eqnarray}
where
\begin{eqnarray}
\tan\phi_{\pm 2}=\frac{-\sin\chi\sin\theta}{\cos\theta\pm\cos\chi}, \ \
\tan\phi_0      =\sin\chi\cot\theta.
\end{eqnarray}
It follows from this or by direct calculation that
\begin{eqnarray}
\phi_2+\phi_{-2}-2\phi_0=\pm\pi.
\label{prel}
\end{eqnarray}
Because of this restriction, the ground state of the cyclic phase is
degenerate with respect to at least two  continuous phase variables.
We therefore expect to have at least two  Goldstone modes, as will be shown in
Sec.\ref{sec:CB}.

\subsection{The phase boundaries}
\label{sec:boundaries}

In the absence of external magnetic field, the ground-state energies
for the three phases are given from Eqs.~(\ref{OP_F}), (\ref{e2}),
and (\ref{cyclic}) by
\begin{eqnarray}
\epsilon^{\rm F}_0=4c_1, \ \
\epsilon^{\rm AF}_0=\frac{c_2}{5}, \ \
\epsilon^{\rm C}_0=0.
\label{H=0}
\end{eqnarray}
It follows that each phase is specified by~\cite{Koashi,Ciobanu}.
\begin{eqnarray}
& {\rm ferromagnetic}     & \ c_1<0 \ {\rm and} \ c_1-c_2/20<0, \\
& {\rm antiferromagnetic} & \ c_2<0 \ {\rm and} \ c_1-c_2/20>0, \\
& {\rm cyclic}            & \ c_1>0 \ {\rm and} \ c_2>0.
\label{PB1}
\end{eqnarray}

In the presence of external magnetic field we define each phase as follows:
\begin{eqnarray*}
& {\rm ferromagnetic}     & \ \langle\hat{f}_z\rangle=2,    \\
& {\rm antiferromagnetic} & \ \langle\hat{s}_-\rangle\neq0, \\
& {\rm cyclic}            & \ \langle\hat{s}_-\rangle=0 \ {\rm and} \
\langle\hat{f}_z\rangle<2.
\end{eqnarray*}
By directly comparing the energies in Eqs.~(\ref{e2}), (\ref{OP_F}),
and (\ref{cyclic}), we find that each phase is specified by the
following conditions:
\begin{eqnarray}
& {\rm ferromagnetic}
          & \ c_1\leq\tilde{p}/4 \ {\rm and} \ c_1-c_2/20\leq\tilde{p}/4,
\label{condF}
\\
& {\rm antiferromagnetic}
          & \ c_2<0 \ {\rm and} \ c_1-c_2/20>\tilde{p}/4,
\label{condAF}
\\
& {\rm cyclic}
          & \ c_1>\tilde{p}/4 \ {\rm and} \ c_2>0.
\label{condC}
\end{eqnarray}

\section{Many-body spin correlations and magnetic response}
\label{sec:MBT}

In this section we study the case in which the system is 
so tightly confined that the coordinate part of the ground-state 
wave function $\phi_0({\bbox r})$ is independent of the spin state 
and solely determined by $\hat{H}_{\rm KE}$, $\hat{H}_{\rm PE}$, and 
the spin-independent part of $\hat{V}$; that is,
$\phi_0({\bbox r})$ is the solution $\phi$ to the equation
\begin{eqnarray}
\left[-\frac{\hbar^2\bbox{\nabla}^2}{2M}+U_{\rm trap}+
c_0(N-1)|\phi|^2\right]\phi=\epsilon\phi
\end{eqnarray}
with the lowest eigenvalue $\epsilon=\epsilon_0$.
This assumption is justified if the second lowest eigenvalue
$\epsilon_1$ satisfies
\begin{eqnarray}
\epsilon_1-\epsilon_0\gg |p|,|c_1|N/V^{\rm eff},|c_2|N/V^{\rm eff},
\label{epsi}
\end{eqnarray}
where $V^{\rm eff}\equiv(\int d\bbox{r}|\phi_0|^4)^{-1}$ 
is an effective volume which coincides with $V$ in Eq.~(\ref{PWE}) 
for the spatially uniform case (i.e., $U_{\rm trap}=0$).
When the condition (\ref{epsi}) is met, 
the field operator $\hat{\Psi}_m$ may be approximated as
$\hat{\Psi}_m\simeq\hat{a}_m\phi_0$, where $\hat{a}_m$ is the annihilation
operator of the bosons that are specified by the spin component $m$ 
and by the coordinate wave function $\phi_0$.
The spin-dependent part of the Hamiltonian can then be written as
\begin{equation}
\hat{H}=
\frac{c_1}{2V^{\rm eff}}:\hat{\bbox{\cal F}}^2:
+\frac{2c_2}{5V^{\rm eff}}\hat{\cal S}_+\hat{\cal S}_-
-p\hat{\cal F}_z,
\label{h2}
\end{equation}
where
\begin{eqnarray}
\hat{\bbox{\cal F}}\equiv\sum_{mn}\bbox{f}_{mn}\hat{a}_m^\dagger\hat{a}_n, \
\
\hat{\cal S}_-=(\hat{\cal S}_+)^\dagger\equiv
\frac{1}{2}\sum_m(-1)^m\hat{a}_m\hat{a}_{-m},
\ \
\hat{\cal F}_z\equiv\sum_{m}m\hat{a}_m^\dagger\hat{a}_m.
\label{FSF}
\end{eqnarray}

\subsection{Spectrum and degeneracy}
\label{sec:Spectrum and degeneracy}

We first make some remarks on the properties of the operators
$\hat{\cal S}_-=\hat{\cal S}_+^\dagger=
(\hat{a}_0)^2/2
-\hat{a}_1\hat{a}_{-1}
+\hat{a}_2\hat{a}_{-2}$.
The operator $\hat{\cal S}_+$, when applied to the vacuum, creates
a pair of bosons in the spin-singlet state. This pair, however,
should not be regarded as a single composite boson
because $\hat{\cal S}_+$ does not satisfy the
commutation relations of bosons.
The operator $\hat{\cal S}_+$ instead
satisfies the $SU(1,1)$ commutation relations together with
$\hat{\cal S}_z\equiv (2\hat{N}+5)/4$, namely,
\begin{eqnarray}
[\hat{\cal S}_z,\hat{\cal S}_\pm]=\pm\hat{\cal S}_\pm, \ \
[\hat{\cal S}_+,\hat{\cal S}_-]=-2\hat{\cal S}_z,
\end{eqnarray}
where the minus sign in the last equation is the only distinction
from the usual spin commutation relations. As a consequence, the
Casimir operator $\hat{\cal S}^{2}$ that commutes with
$\hat{\cal S}_\pm$ and $\hat{\cal S}_z$ is given by
\begin{eqnarray}
\hat{\cal S}^{2}\equiv -\hat{\cal S}_+\hat{\cal S}_-
+\hat{\cal S}_z^2-\hat{\cal S}_z.
\end{eqnarray}

Consider an eigenspace ${\cal H}_\nu$
of $\hat{\cal S}^{2}$ with an eigenvalue $\nu$.
The requirement that
$\hat{\cal S}_+\hat{\cal S}_-=\hat{\cal S}_z^2-\hat{\cal S}_z
-\hat{\cal S}^{2}$ must be positive semidefinite
means that, in ${\cal H}_\nu$, the eigenvalues of
the operator $\hat{\cal S}_z$ has a minimum value
$(2N_0+5)/4$, where $N_0$ is a nonnegative integer.
For a state $|\phi\rangle$ that belongs to the minimum eigenvalue,
the norm of $\hat{\cal S}_-|\phi\rangle$ must vanish;
hence $\nu=S(S-1)$ with $S=(2N_0+5)/4$.
We thus obtain the allowed combinations of eigenvalues
$\{S(S-1),S_z\}$ for $\hat{\cal S}^{2}$
and $\hat{\cal S}_z$ such that
\begin{equation}
S=(2N_0+5)/4 \;\;
(N_0=0,1,2,\ldots)
\label{eigen_S}
\end{equation}
    and
\begin{equation}
S_z=S+N_{\rm S} \;\;
(N_{\rm S}=0,1,2,\ldots).
\label{def_N2}
\end{equation}
Here we have introduced
quantum numbers $N_{\rm S}$ and $N_0$, where the operator
$\hat{\cal S}_+$ raises $N_{\rm S}$ by one and the relation
\begin{equation}
N=2N_{\rm S}+N_0
\end{equation}
holds. We  may thus interpret $N_{\rm S}$ as the number of
spin-singlet `pairs', and $N_0$ as that of the remaining bosons.

Exact energy eigenvalues of Hamiltonian (\ref{h2}) can be obtained
as follows.
The operators $\hat{{\cal S}}_\pm$ are invariant
under any rotation of the system, namely,
they commute with the total spin operator $\hat{\bbox{\cal F}}$.
The energy eigenstates can thus be classified
according to quantum numbers
$N_0$ and
$N_{\rm S}$,
total spin $F$,
and magnetic quantum number $F_z$.
We thus denote the eigenstates as
$|N_0,N_{\rm S},F,F_z;\lambda\rangle$, where
$\lambda=1,2,\ldots,g_{N_0,F}$ labels orthonormal
degenerate states, that is,
\begin{eqnarray}
\langle N_0,N_{\rm S},F,F_z;\lambda^\prime|
N_0,N_{\rm S},F,F_z;\lambda\rangle=
\delta_{\lambda^\prime,\lambda}.
\end{eqnarray}
The number of degenerate states $g_{N_0,F}$ for a given set of
$\{N_0,N_{\rm S},F,F_z\}$ will be referred to as the size of
the eigenspaces for $\{N_0,N_{\rm S},F,F_z\}$.
It will be shown to be independent of $N_{\rm S}$ and $F_z$ below.
The energy eigenvalue for the state $|N_0,N_{\rm S},F,F_z;\lambda\rangle$
is given by
\begin{equation}
E=\frac{c_1}{2V^{\rm eff}}[F(F+1)-6N]
+\frac{c_2}{5V^{\rm eff}}N_{\rm S}(N+N_0+3)
-pF_z,
\label{E2}
\end{equation}
where the relation $2N_{\rm S}+N_0=N$ is used.

The degeneracy $g_{N_0,F}$ can be calculated
as follows. First we show that $g_{N_0,F}$ is independent of
$N_{\rm S}$ and $F_z$. This is seen by the following relations,
\begin{equation}
\langle N_0,N_{\rm S},F,F_z;\lambda^\prime|
\hat{\cal F}_\pm\hat{\cal F}_\mp
|N_0,N_{\rm S},F,F_z;\lambda\rangle=
(F\pm F_z)(F\mp F_z+1)\delta_{\lambda^\prime,\lambda},
\end{equation}
where $\hat{\cal F}_\pm\equiv\hat{\cal F}_x\pm i\hat{\cal F}_y$,
\begin{equation}
\langle N_0,N_{\rm S},F,F_z;\lambda^\prime|\hat{{\cal S}}_+\hat{{\cal
S}}_- |N_0,N_{\rm S},F,F_z;\lambda\rangle=
N_{\rm S}(N_0+N_{\rm S}+3/2)\delta_{\lambda^\prime,\lambda},
\end{equation}
and
\begin{equation}
\langle N_0,N_{\rm S},F,F_z;\lambda^\prime|\hat{{\cal S}}_-\hat{{\cal
S}}_+ |N_0,N_{\rm S},F,F_z;\lambda\rangle=
(N_{\rm S}+1)(N_0+N_{\rm S}+5/2)\delta_{\lambda^\prime,\lambda}.
\label{formula_S_plus}
\end{equation}
These relations implies that the sizes of the eigenspaces
for $\{N_0,N_{\rm S}\pm 1,F,F_z\pm 1\}$ are not smaller than the size of
the eigenspace for $\{N_0,N_{\rm S},F,F_z\}$. The degeneracy thus depends
only on $N_0$ and $F$. Next, we introduce a generating function of
$g_{N_0,F}$ defined by
\begin{equation}
G(x,y)\equiv\sum_{N_0=0}^\infty \sum_{F=0}^\infty g_{N_0,F}x^{N_0}y^F.
\label{gf_def}
\end{equation}
Let $h_{N,F_z}$ be the total number of states with a fixed number of
bosons $N$ and a fixed magnetic quantum number $F_z$.
This is given by the total number of
combinations of nonnegative integers
$\{n_{-2},n_{-1},n_{0},n_{1},n_{2}\}$ that satisfy
$n_{-2}+n_{-1}+n_{0}+n_{1}+n_{2}=N$ and
$-2n_{-2}-n_{-1}+n_{1}+2n_{2}=F_z$. 
It follows that 
\begin{equation}
\sum_{N=0}^\infty \sum_{F_z=-\infty}^{\infty}
h_{N,F_z} z^N y^{2N+F_z+1}=
\sum_{\{n_j\}} \prod_{j=-2}^2 z^{n_j} y^{ (j+2) n_j+1}
=y\prod_{j=-2}^2 (1-zy^{j+2})^{-1},
\label{gf_h}
\end{equation}
where we assume $|y|<1$ and $|z|<1$ to ensure the convergence of the series.
Let $\tilde{h}_{N,F,F_z}$ be the total number of states 
for given $N$, $F$, and $F_z$.
Because $\tilde{h}_{N,F,F_z}$ is independent of the value of $F_z$, we shall
denote it simply as $\tilde{h}_{N,F}$.

The quantity $h_{N,F_z}$ is written in terms of the sum of
$\tilde{h}_{N,F}$ as
\begin{equation}
h_{N,F_z}
=\sum_{F(\ge |F_z|)} \tilde{h}_{N,F},
\end{equation}
and hence $\tilde{h}_{N,F}=h_{N,F}-h_{N,F+1}$. Let us extend the
definition of $\tilde{h}_{N,F}$ to the negative values of $F$
through this relation. It follows then from Eq.~(\ref{gf_h}) that
\begin{equation}
\sum_{N=0}^\infty \sum_{F=-\infty}^{\infty}
\tilde{h}_{N,F} z^N y^{2N+F+1}
=(y-1)\prod_{j=-2}^2 (1-zy^{j+2})^{-1}.
\end{equation}
The right hand side of this equation can be written as
the sum of two fractions $G_1(z,y)+G_2(z,y)$, where
    \begin{equation}
G_1(z,y)
=\frac{y(1-zy^3+z^2y^6)}{(1-zy^3)(1-zy^4)(1-z^2y^4)(1-z^3y^6)}
\label{def_gf1}
\end{equation}
and
    \begin{equation}
G_2(z,y)
=-\frac{1-zy+z^2y^2}{(1-z)(1-zy)(1-z^2y^4)(1-z^3y^6)}.
\end{equation}
Making Maclaurin expansions of $G_1$ and $G_2$ around
$z=y=0$ and regrouping them in terms of the form $z^{N}y^{2N+F+1}$,
we find that $G_1$ consists only of the terms with $F\ge 0$, and
$G_2$ of those with $F<0$. We thus obtain
\begin{equation}
\sum_{N=0}^\infty \sum_{F=0}^{\infty}
\tilde{h}_{N,F} z^{N} y^{2N+F+1}
=G_1(z,y).
\label{gf_tildeh}
\end{equation}
The quantity $\tilde{h}_{N,F}$ is written by the sum of the degeneracy
$g_{N_0,F}$ as
\begin{equation}
\tilde{h}_{N,F}
=\sum_{N_{\rm S}(\le N/2)} g_{N-2N_{\rm S},F},
\end{equation}
and hence we can write
$g_{N_0,F}=\tilde{h}_{N_0,F}-\tilde{h}_{N_0-2,F}$,
where we assume that $\tilde{h}_{-1,F}=\tilde{h}_{-2,F}=0$.
It follows then from Eq.~(\ref{gf_tildeh}) that
\begin{equation}
\sum_{N_0=0}^\infty \sum_{F=-\infty}^{\infty}
g_{N_0,F} z^{N_0} y^{2N_0+F+1}
=(1-z^2y^4)G_1(z,y),
\label{gf_g}
\end{equation}
and we finally obtain an explicit form of the generating function
$G(x,y)$ defined by Eq.~(\ref{gf_def}) as
\begin{equation}
\sum_{N_0=0}^\infty \sum_{F=0}^\infty
g_{N_0,F}x^{N_0}y^F=y^{-1}(1-x^2)G_1(xy^{-2},y)
=\frac{1-xy+x^2y^2}{(1-xy)(1-xy^2)(1-x^3)}.
\label{def_gf}
\end{equation}
The total spin $F$ can, in general, take integer values in the range
$0\le F \le 2N_0$. However, from Eq.~(\ref{def_gf}) we find that there
are some forbidden values. That is, $F=1,2,5,2N_0-1$ are not allowed
when $N_0=3k(k\in \bbox{Z})$, and  $F=0,1,3,2N_0-1$ are forbidden when
$N_0=3k\pm1$.
An easier way to find the forbidden values is discussed at the end of 
Sec.~\ref{sec:Energy Eigenstates}.

\subsection{Energy eigenstates}
\label{sec:Energy Eigenstates}

The energy eigenstates $|N_0,N_{\rm S},F,F_z;\lambda\rangle$ can be 
constructed using one-, two- and three-boson creation operators.
Let us define the operator $\hat{A}^{(n)}_f{}^\dagger$ such that it
creates $n$ bosons in the state with total spin $F=f$ and magnetic
quantum number $F_z=f$ when applied to the vacuum.
Such states are unique when $n\le 3$. 
Among possible operators $\hat{A}^{(n)\dagger}_f$, 
we choose the following five operators for constructing the eigenstates:
\begin{eqnarray}
\hat{A}^{(1)\dagger}_2&=&\hat{a}_2{}^\dagger
\label{def_A12}
\\
\hat{A}^{(2)\dagger}_0&=&
\frac{1}{\sqrt{10}}
[(\hat{a}_0^\dagger)^2-2\hat{a}_1^\dagger\hat{a}_{-1}^\dagger
+2\hat{a}_2^\dagger\hat{a}_{-2}^\dagger]
=\sqrt{\frac{2}{5}}\hat{\cal S}_+
\\
\hat{A}^{(2)\dagger}_2&=&
\frac{1}{\sqrt{14}}
[2\sqrt{2}\hat{a}_2^\dagger\hat{a}_0^\dagger
-\sqrt{3}(\hat{a}_1^\dagger)^2]
\\
\hat{A}^{(3)\dagger}_0&=&
\frac{1}{\sqrt{210}}
[\sqrt{2}(\hat{a}_0^\dagger)^3-3\sqrt{2}\hat{a}_1\dagger
\hat{a}_0^\dagger\hat{a}_{-1}^\dagger
+3\sqrt{3}(\hat{a}_1^\dagger)^2\hat{a}_{-2}^\dagger
+3\sqrt{3}\hat{a}_2^\dagger(\hat{a}_{-1}^\dagger)^2
-6\sqrt{2}\hat{a}_2^\dagger\hat{a}_0^\dagger
\hat{a}_{-2}^\dagger]
\label{def_A30}
\\
\hat{A}^{(3)\dagger}_3&=&
\frac{1}{\sqrt{20}}
[(\hat{a}_1^\dagger)^3-\sqrt{6}\hat{a}_2^\dagger\hat{a}_1^\dagger
\hat{a}_0^\dagger
+2(\hat{a}_2^\dagger)^2\hat{a}_{-1}^\dagger].
\end{eqnarray}
Note that $\hat{A}^{(2)\dagger}_1$ and $\hat{A}^{(3)\dagger}_1$ do not
exist because of the Bose symmetry.
Note also that the operators $\hat{A}^{(n)\dagger}_f$ commute with $\hat{F}_+$.

Consider a set ${\cal B}$ of unnormalized states,
\begin{equation}
|n_{12},n_{20},n_{22},n_{30},n_{33}\rangle
\equiv
(\hat{a}^\dagger_2)^{n_{12}}
(\hat{A}^{(2)\dagger}_0)^{n_{20}}
(\hat{A}^{(2)\dagger}_2)^{n_{22}}
(\hat{A}^{(3)\dagger}_0)^{n_{30}}
(\hat{A}^{(3)\dagger}_3)^{n_{33}}
|{\rm vac}\rangle
\label{def_state_unit}
\end{equation}
with $n_{12},n_{20},n_{22},n_{30}=0,1,2,\ldots,\infty$ and $n_{33}=0,1$. 
The state
$|n_{12},n_{20},n_{22},n_{30},n_{33}\rangle$
has the total number of bosons
$N=n_{12}+2(n_{20}+n_{22})+3(n_{30}+n_{33})$
and the total spin
$F=F_z=2(n_{12}+n_{22})+3n_{33}$.
When $N$ and $F$ are given, $n_{33}$ is uniquely
determined through the parity of $F$, namely,
\begin{eqnarray}
n_{33}=F\;{\rm mod}\; 2.
\end{eqnarray}
If we introduce the following two parameters
\begin{eqnarray}
\mu&\equiv & 2n_{20}+3n_{30},
\\
\nu&\equiv & n_{20}+n_{30},
\end{eqnarray}
$n_{20}$ and $n_{30}$ are uniquely specified by them, 
and the remaining $n_{12}$ and $n_{22}$ are also determined as
\begin{eqnarray}
n_{12}&= & F-N+\mu,
\\
n_{22}&= & N-\frac{F}{2}-\mu-\frac{3}{2}n_{33}.
\end{eqnarray}
The parameter set $\{n_{12},n_{20},n_{22},n_{30},n_{33}\}$
is thus uniquely specified by the set $\{N, F, \mu, \nu\}$.
Let us consider the two commuting observables
\begin{eqnarray}
\hat{n}_\mu&\equiv & 2\hat{a}_{-2}^\dagger\hat{a}_{-2}
+\frac{3}{2}(\hat{a}_{-1}^\dagger\hat{a}_{-1}-\hat{n}_{33})
\\
\hat{n}_\nu&\equiv & \hat{a}_{-2}^\dagger\hat{a}_{-2}
+\frac{1}{2}(\hat{a}_{-1}^\dagger\hat{a}_{-1}-\hat{n}_{33})
\end{eqnarray}
with $\hat{n}_{33}\equiv \hat{F}\;{\rm mod}\; 2$.
Let $\hat{P}(n_\mu, n_\nu)$ be the projection operator
onto the simultaneous eigenspace of $\hat{n}_\mu$ and $\hat{n}_\nu$
corresponding to eigenvalues $n_\mu$ and $n_\nu$, respectively.
Noting that $\hat{A}^{(2)}_0$ includes the term
$\hat{a}_2\hat{a}_{-2}$ and $\hat{A}^{(3)}_0$ includes
$\hat{a}_2(\hat{a}_{-1})^2$,
it can be seen from Eqs.~(\ref{def_A12})-(\ref{def_state_unit}) that
\begin{equation}
\hat{P}(n_\mu, n_\nu)|n_{12},n_{20},n_{22},n_{30},n_{33}\rangle
\neq \bbox{0} \;\;{\rm if} \;\; n_\mu= \mu \;{\rm and}\;
n_\nu=\nu,
\end{equation}
and also
\begin{equation}
\hat{P}(n_\mu, n_\nu)|n_{12},n_{20},n_{22},n_{30},n_{33}\rangle
= \bbox{0} \;\;{\rm if} \;\; n_\mu> \mu \;{\rm or}\;
n_\nu>\nu.
\end{equation}
If we order the pair $(\mu,\nu)$ in the lexicographic way,
namely, $(\mu,\nu)<(\mu',\nu')$ if
$\mu<\mu'$ or $(\mu=\mu'$ and $\nu<\nu')$,
the above equations imply that a state in ${\cal B}$ specified by
$\{N, F, \mu, \nu\}$ is linearly independent of the set of
states specified by $\{N, F, \mu', \nu'\}$ with
$(\mu',\nu')<(\mu,\nu)$. The set  ${\cal B}$ is
thus a linearly independent set of states.
Let $\tilde{h}'_{N,F}$ be the total number of states in ${\cal B}$ with
$N$ bosons, total spin $F$, and magnetic quantum number $F_z=F$.
A generating function of $\tilde{h}'_{N,F}$ is calculated as
\begin{eqnarray}
\sum_{N=0}^\infty \sum_{F=0}^{\infty}
\tilde{h}'_{N,F} x^{N} y^{F}
&=&\sum_{n_{12},n_{20},n_{22},n_{30}=0}^\infty
\sum_{n_{33}=0}^1
x^{n_{12}+2(n_{20}+n_{22})+3(n_{30}+n_{33})}
y^{2(n_{12}+n_{22})+3n_{33}}
\nonumber \\
&=&\frac{1+x^3y^3}{(1-xy^2)(1-x^2)(1-x^2y^2)(1-x^3)}
=y^{-1}G_1(xy^{-2},y),
\end{eqnarray}
where $G_1$ is defined by Eq.~(\ref{def_gf1}).
Compared to Eq.~(\ref{gf_tildeh}), we have
$\tilde{h}'_{N,F}=\tilde{h}_{N,F}$.
This implies that ${\cal B}$ is complete,
namely, the set ${\cal B}$ forms a nonorthogonal basis
of the subspace ${\cal H}_{(F_z=F)}$
in which magnetic quantum number $F_z$
is equal to total spin $F$.

The energy eigenstates can be obtained by partially applying the
method of Schmidt's orthogonalization to the nonorthogonal basis ${\cal B}$.
Here by ^^ ^^ partially" we mean that eigenstates corresponding to different
energies are orthogonal, but that those corresponding to the same energy are
not always so.
Let us consider a series of subspaces
${\cal H}_{(F_z=F)}=
{\cal H}^{(0)}\supset{\cal H}^{(1)}\supset\cdots$,
where
${\cal H}^{(j)}$ is spanned by the states with quantum number $N_{\rm S}$
[see Eq.~(\ref{def_N2})] satisfying $N_{\rm S}\ge j$.
Let $\hat{P}^{(j)}$ be the projection operator onto ${\cal H}^{(j)}$.
From the relation 
$\hat{\cal S}_+ \hat{P}^{(j)}=\hat{P}^{(j+1)}\hat{\cal S}_+$,
we have $\hat{P}^{(j)}(\hat{A}^{(2)\dagger}_0)^j
=(\hat{A}^{(2)\dagger}_0)^j\hat{P}^{(0)}$.
Since $\hat{P}^{(0)}|\phi\rangle=|\phi\rangle$ for any state
$|\phi\rangle\in {\cal B}$,
we have
\begin{equation}
(\hat{P}^{(0)}-\hat{P}^{(n_{20})})|n_{12},n_{20},n_{22},n_{30},n_{33}\rangle
= \bbox{0},
\end{equation}
implying that ${\cal H}^{(j)}$ is spanned by all the states
$\{|n_{12},n_{20},n_{22},n_{30},n_{33}\rangle\}$ satisfying
$n_{20}\ge j$. We can then construct a new basis ${\cal B}'$
made up of the states of the form
\begin{equation}
(\hat{P}^{(0)}-\hat{P}^{(n_{20}+1)})|n_{12},n_{20},n_{22},n_{30},n_{33}
\rangle
=(\hat{A}^{(2)\dagger}_0)^{n_{20}}
\hat{P}_{(N_{\rm S}=0)}
(\hat{a}^\dagger_2)^{n_{12}}
(\hat{A}^{(2)\dagger}_2)^{n_{22}}
(\hat{A}^{(3)\dagger}_0)^{n_{30}}
(\hat{A}^{(3)\dagger}_3)^{n_{33}}
|{\rm vac}\rangle,
\end{equation}
where $\hat{P}_{(N_{\rm S}=0)}\equiv \hat{P}^{(0)}-\hat{P}^{(1)}$
is the projection onto the subspace with $N_{\rm S}=0$
(the kernel of $\hat{\cal S}_-$).
It is easy to see that the states belonging to ${\cal B}'$
are simultaneous eigenstates of $\{\hat{N}, \hat{\cal S}_+\hat{\cal S}_-,
\hat{\bbox{\cal F}}^2, \hat{\cal F}_z\}$, and hence energy eigenstates.
The energy eigenstates with $F_z< F$ can be constructed by applying
$(\hat{\cal F}_-)^{F-F_z}$ to the states of ${\cal B}'$.

To summarize, the energy eigenstates can be represented as
\begin{equation}
(\hat{\cal F}_-)^{\Delta F}(\hat{A}^{(2)\dagger}_0)^{n_{20}}
\hat{P}_{(N_{\rm S}=0)}
(\hat{a}^\dagger_2)^{n_{12}}
(\hat{A}^{(2)\dagger}_2)^{n_{22}}
(\hat{A}^{(3)\dagger}_0)^{n_{30}}
(\hat{A}^{(3)\dagger}_3)^{n_{33}}
|{\rm vac}\rangle,
\label{eigenstates}
\end{equation}
with $n_{12},n_{20},n_{22},n_{30}=0,1,2,\ldots,\infty$,
$n_{33}=0,1$, and $\Delta F=0,1,\ldots, 2F$.
These parameters are related to $\{N_0,N_{\rm S},F,F_z\}$ as
\begin{eqnarray}
N_0&=&n_{12}+2n_{22}+3n_{30}+3n_{33}
\\
N_{\rm S}&=&n_{20}
\\
F&=&2n_{12}+2n_{22}+3n_{33}
\\
F_z&=&F-\Delta F,
\end{eqnarray}
and the corresponding eigenenergy is given by Eq.~(\ref{E2}).
Note that the states defined in (\ref{eigenstates})
are unnormalized, and the states having the same energy 
(i.e., those belonging to the same set of parameter values 
$\{N_0,N_{\rm S},F,F_z\}$) are nonorthogonal.

The representation (\ref{eigenstates}) of the energy eigenstates utilizes 
the operator $\hat{A}^{(n)\dagger}_f$ that creates correlated $n$   
bosons having total spin $f$. 
It might be tempting to envisage a physical picture that the system is, 
like in $^4$He, made up of $n_{nf}$ composite bosons whose creation operator 
is given by $\hat{A}^{(n)\dagger}_f$. 
However, this picture is oversimplified.
First of all, the operator $\hat{A}^{(n)}_f$ does not obey the boson 
commutation relation.
In addition, the projection operator $\hat{P}_{(N_{\rm S}=0)}$ in
(\ref{eigenstates}) imposes many-body spin correlations such that
the spin correlation between {\it any} two bosons must have
vanishing spin-singlet component. 
Note that two bosons with independently fluctuating spins have
a nonzero overlap with the spin-singlet state in general.
The many-body spin correlations of the energy eigenstates are 
thus far more complicated than what an intuitive picture of composite 
bosons suggests. 
On the other hand, as long as quantities such as the number of bosons, 
magnetization, and energy are concerned, 
the above simplified picture is quite helpful. 
By way of illustration, we provide an alternative explanation for 
the existence of forbidden values for the total spin $F$, which were
found earlier using the generating function (\ref{def_gf}).
For example, to construct a state with $F=0$ or $F=3$,
composite particles with total spin 2 must be avoided, namely,
$n_{12}=n_{22}=0$. Then we have $N_0=3(n_{30}+n_{33})$, implying that
$F=0$ or $F=3$ is only possible when $N_0=3k(k\in \bbox{Z})$.
For a state with $F=2$ or $F=5$, we have $n_{12}+n_{22}=1$ and
$N_0=1+n_{22}+3(n_{30}+n_{33})$ implying that $N_0\neq 3k(k\in \bbox{Z})$.
The above simplified picture is also helpful when we consider the magnetic
response as discussed below.

\subsection{Magnetic response}
\label{sec:Magnetic response}

We consider here how the ground state and the magnetization $F_z$ 
respond to the applied magnetic field $p$.
From Eq.~(\ref{E2}), we see that the minimum energy states always 
satisfy $F_z=F$ when $p>0$. The problem thus reduces to minimizing
the function
\begin{equation}
E(F_z,N_s)=\frac{c_1}{2V^{\rm eff}}[F_z(F_z+1)-6N]
+\frac{c_2}{5V^{\rm eff}}N_{\rm S}(2N-2N_{\rm S}+3)
-pF_z.
\label{E3}
\end{equation}
For this purpose, it is convenient to consider the cases $c_2>0$ and 
$c_2<0$ separately.

\subsubsection{Case of $c_2>0$}

Let us first consider the case $c_2>0$.

When $c_1< 0$, the energy (\ref{E3}) is minimized when 
$N_{\rm S}=0$, $N_0=N$, and $F=F_z=2N$, and 
the ground state is given by 
$(\hat{a}^\dagger_2)^{N}|{\rm vac}\rangle$, 
that is, the system is ferromagnetic.
This result is consistent with that obtained from MFT.

When $c_1>0$, let us rewrite the energy as
\begin{equation}
E(F_z,N_s)=\frac{c_1}{2V^{\rm eff}}
\left(F_z-\frac{pV^{\rm eff}}{c_1}+\frac{1}{2}\right)^2
-\frac{2c_2}{5V^{\rm eff}}
\left(N_{\rm S}-\frac{2N+3}{4}\right)^2+{\rm const.}
\end{equation}
The energy is thus lower when $F_z$ is closer to $pV^{\rm eff}/c_1-1/2$ and
when $N_{\rm S}$ is smaller. In most of the parameter space,
the ground state is
$|N_{\rm S}=0,N_0=N,F,F_z=F;\lambda\rangle$ with $F$
taking the allowed integer closest to $pV^{\rm eff}/c_1-1/2$.
Since these states satisfy $\langle \hat{S}_+ \hat{S}_-\rangle=0$,
they belong to the ferromagnetic phase or the cyclic phase.
The separatrix between the two phases is given by
$pV^{\rm eff}/c_1-1/2=2N-1$.
We thus find that the ground state is ferromagnetic if
$c_1<pV^{\rm eff}/(2N-1/2)\simeq pV^{\rm eff}/(2N)$ and cyclic otherwise.
This classification is consistent with the mean-field analysis given 
in Sec.~\ref{sec:boundaries}.
As seen in Sec.~\ref{sec:Spectrum and degeneracy},
the above ground state is highly degenerated in general;
this may originate from the continuous symmetry that leaves free
at least two  parameters characterizing the order parameter of the cyclic phase
as shown in Sec.~\ref{sec:CBEC}.
According to the discussions in Sec.~\ref{sec:Spectrum and degeneracy}, 
the degeneracy of the states
$|N_{\rm S}=0,N_0=N,F,F_z=F;\lambda\rangle$ is equal to the number of the
combinations of $\{n_{12},n_{22},n_{30}\}$ satisfying
$n_{12}+2n_{22}+3n_{30}=N-3n_{33}$ and
$2n_{12}+2n_{22}=F-3n_{33}$.
The number of trios, $n_{30}$, can take values in the
range $(N-F)/3\lesssim n_{30} \lesssim (N-F/2)/3$.
When the magnetic field is nearly zero and $F\sim 0$,
 there is little degeneracy and $n_{30}\sim N/3$,
namely, almost all bosons form trios.

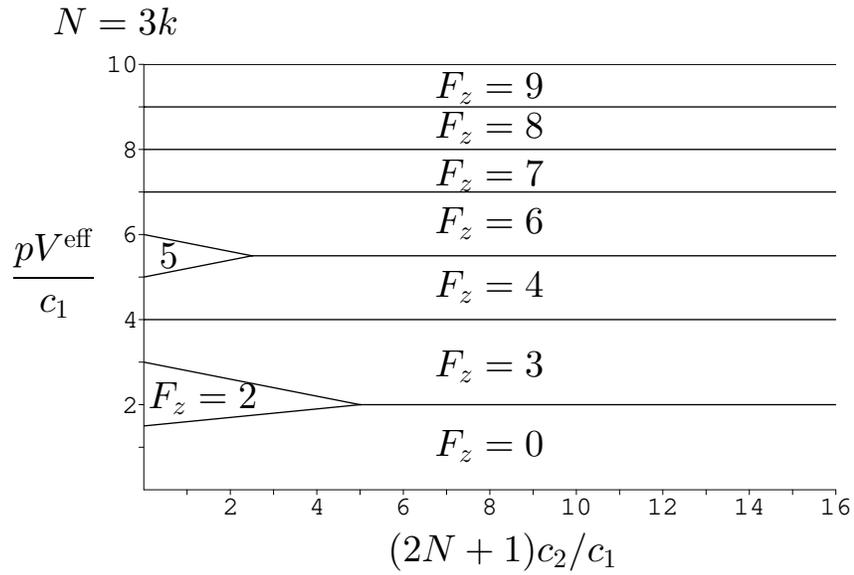
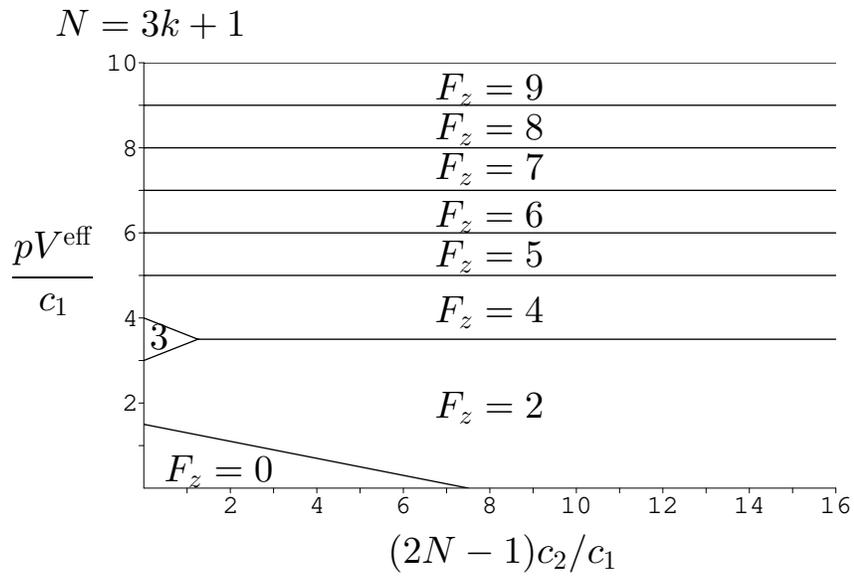
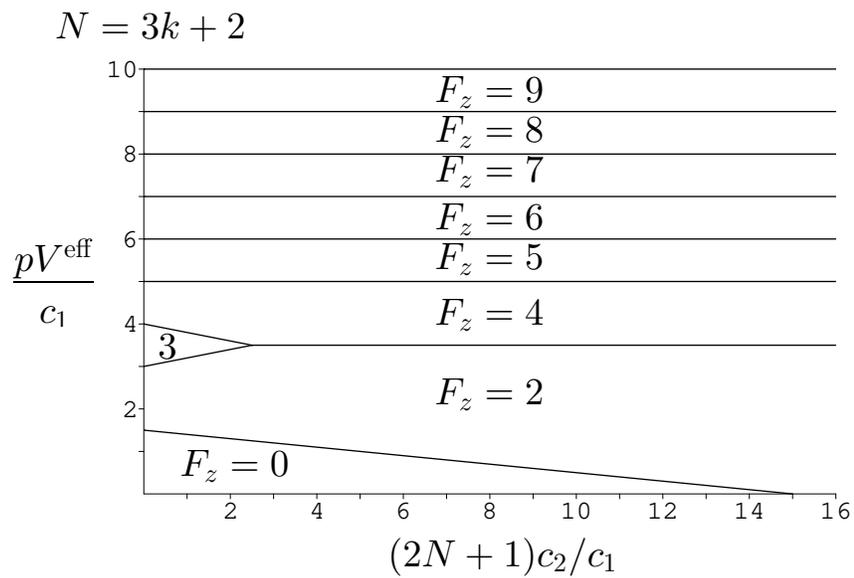
\begin{figure}
\caption{Diagrams of the ground-state magnetization
for $c_2>0$ and $c_1>0$.
\label{fig:1}}
\end{figure}

When $pV^{\rm eff}/c_1-1/2$ is close to the forbidden values of $F$
($F=2,5$ if $N_0=0$ (mod 3), and $F=0,3$ otherwise) for the above state
with $N_{\rm S}=0$, $F_z$ may take those values at the cost of increasing
$N_{\rm S}$ to 1 or 2,
since any of the three values (0,1,2) of $N_0$ mod 3
is realized by setting
$N_{\rm S}$ as 0, 1, or 2 [recall the relation $N=2N_{\rm S}+N_0$].
Whether or not the states $|N_{\rm S}=1,N_0=N-2,F,F_z=F;\lambda\rangle$
and $|N_{\rm S}=2,N_0=N-4,F,F_z=F;\lambda\rangle$ can be the lowest-energy
state depends on the ratio $c_2/c_1$. In Fig~\ref{fig:1}, we give
diagrams of the ground states for small $pV^{\rm eff}/c_1$.

\subsubsection{Case of $c_2<0$}

In this case, it is convenient to introduce a new parameter
\begin{equation}
c_1'\equiv c_1-\frac{c_2}{20},
\end{equation}
and consider the energy as a function of $F_z$ and $l\equiv 2N_0-F_z$:
\begin{equation}
E(F_z,l)=\frac{c_1'}{2V^{\rm
eff}}\left[F_z-\frac{V^{\rm eff}}{c_1'}\left(p+\frac{c_2}{8V^{\rm eff}}
\right)+\frac{1}{2}\right]^2-\frac{c_2}{40V^{\rm eff}}l(l+2F+6)+{\rm const}.
\label{e_fz_l}
\end{equation}
Since $c_2<0$, we see that $E(F_z,l)$ is an increasing function of $l$,
namely,
\begin{equation}
E(F_z,l)>E(F_z,l') \;\; {\rm for} \;\; l>l'.
\end{equation}
Let us consider the cases $c_1'<0$ and  $c_1'>0$ separately.

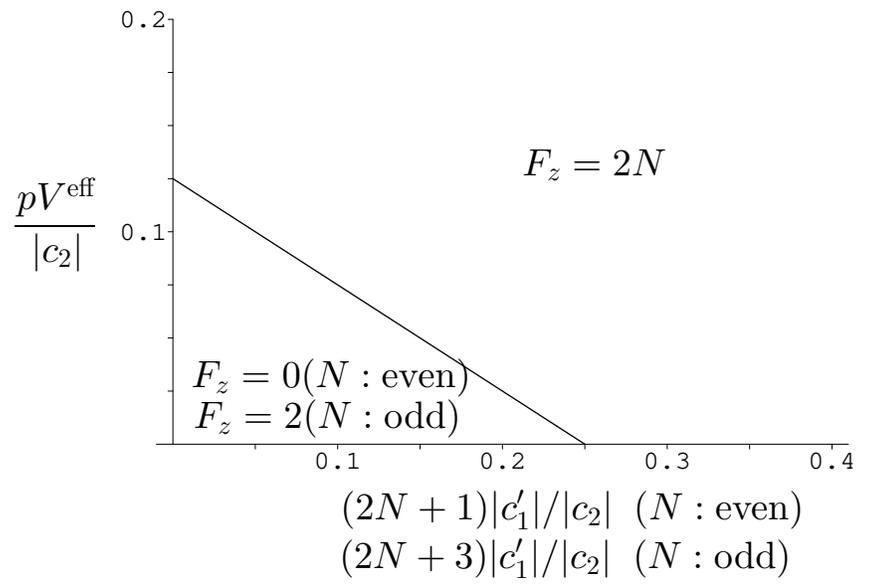
\begin{figure}
\caption{Diagram of the ground-state magnetization
for $c_2<0$ and $c_1'<0$.
\label{fig:2}}
\end{figure}

{\it (a)} $c_1'<0$ ---
In this parameter region, MFT predicts that the system is ferromagnetic,
namely, the lowest-energy state always shows $\langle F_z \rangle=2N$
regardless of the magnitude of magnetic field $p$. In the exact
solution considered here, the magnetic response is different because of
the offset term $c_2/(8V^{\rm eff})$ in Eq.~(\ref{e_fz_l}). Since $c_2<0$,
this term counteracts the applied magnetic field.
It is thus expected that magnetization is suppressed when magnetic field
is weak. The exact ground state is derived as follows. When $N$ is even,
the state $(\hat{A}^{(2)\dagger}_0)^{N/2}|{\rm vac}\rangle$ has
energy $E(F_z=0,l=0)$, and the state
$(\hat{a}_2^{\dagger})^{N}|{\rm vac}\rangle$ has energy
$E(F_z=2N,l=0)$. Any other set $\{{F_z,l}\}$ gives an energy
higher than one of these states.
$(\hat{A}^{(2)\dagger}_0)^{N/2}|{\rm vac}\rangle$ is thus the ground state
when
$E(F_z=0,l=0)<E(F_z=2N,l=0)$, or equivalently,
\begin{equation}
pV^{\rm eff}<c_1'\left(N+\frac{1}{2}\right)-\frac{c_2}{8}
\end{equation}
and otherwise the ground state is
$(\hat{a}_2^{\dagger})^{N}|{\rm vac}\rangle$ [see Fig~\ref{fig:2}].
When $N$ is odd, $F_z=0$ is attained only when $N_0\ge 3$, $F_z=1$ is
forbidden, and the state
$\hat{a}_2^{\dagger}(\hat{A}^{(2)\dagger}_0)^{(N-1)/2}|{\rm vac}\rangle$ has
   energy $E(F_z=2,l=0)$. It is easy to confirm that
$E(F_z=2,l=0)<E(F_z=0,l=6)$ always holds. Therefore,
$\hat{a}_2^{\dagger}(\hat{A}^{(2)\dagger}_0)^{(N-1)/2}|{\rm vac}\rangle$
is the ground state when
$E(F_z=2,l=0)<E(F_z=2N,l=0)$, or equivalently,
\begin{equation}
pV^{\rm eff}<c_1'\left(N+\frac{3}{2}\right)-\frac{c_2}{8},
\end{equation}
and otherwise the ground state is
$(\hat{a}_2^{\dagger})^{N}|{\rm vac}\rangle$ [see Fig~\ref{fig:2}].
These results indicate that in the parameter region of
$c_2\lesssim 8N c_1'<0$,
magnetization of the ground state jumps from 0 or 2 to $2N$.
Such a huge discontinuity does not appear in MFT with a linear
Zeeman effect. (However, in the presence of a quadratic Zeeman
effect, such a jump occurs also in MFT~\cite{Stenger}.)

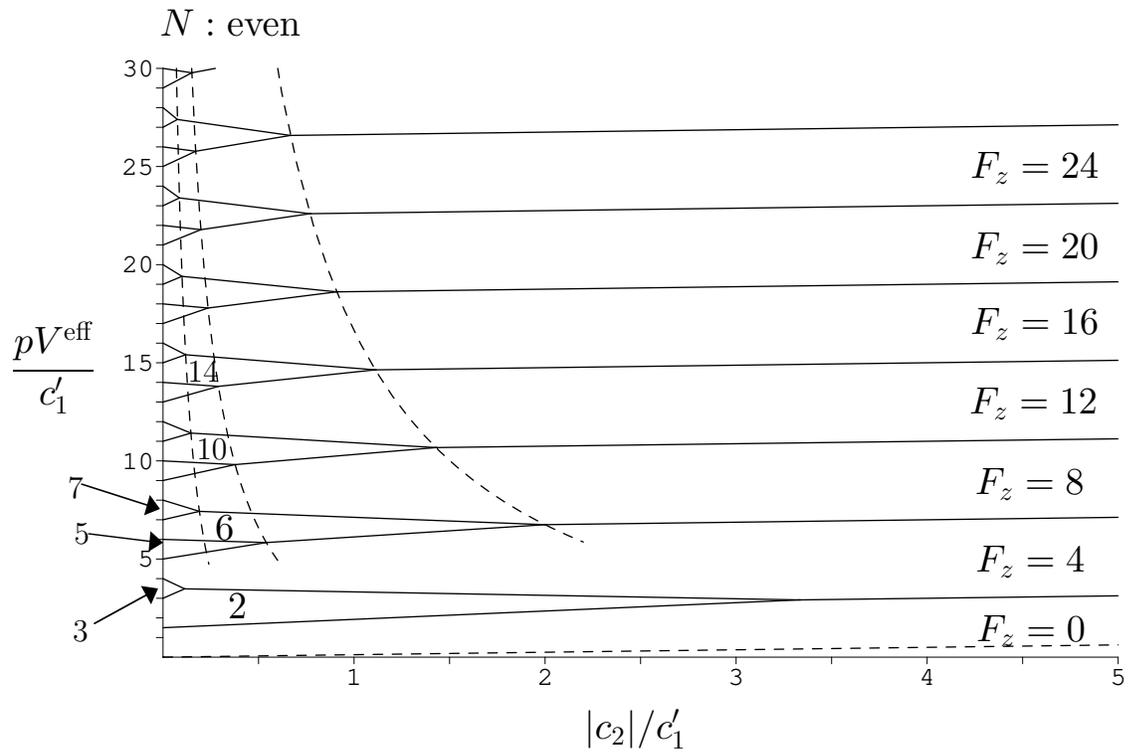
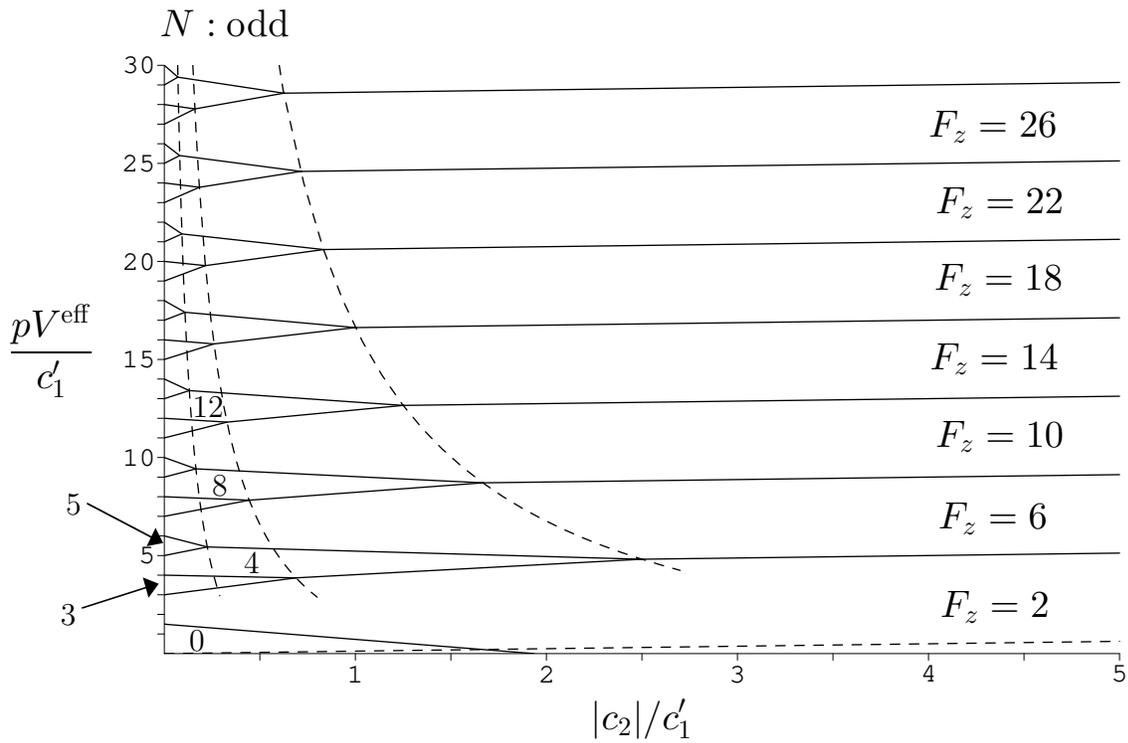
\begin{figure}
\caption{Diagram of the ground-state magnetization
for $c_2<0$ and $c_1'>0$.
\label{fig:3}}
\end{figure}

{\it (b)} $c_1'>0$ --- Given $F_z \ge 6$, the minimum allowed
value of $l$ is determined as follows. Note that
$l=2N-F_z-4N_{\rm S}$ is minimized when the number of singlets $N_{\rm
S}$ is maximized.  For $F_z=2N-4k$ ($k$ is an
integer), the state $(\hat{A}^{(2)\dagger}_0)^{k}
(\hat{a}^\dagger_2)^{N-2k}|{\rm vac}\rangle$ gives $l=0$. To increase
$F_z$ by one ($F_z=2N-4k+1$), one singlet pair must be broken and the
minimum of $l$  is $l=3$ given by the state
$(\hat{A}^{(2)\dagger}_0)^{k-1}
(\hat{a}^\dagger_2)^{N-2k-1}
\hat{A}^{(3)\dagger}_3|{\rm vac}\rangle$.
Keeping the singlet part, $F_z$ can be further increased to
$F_z=2N-4k+2$ by the state $(\hat{A}^{(2)\dagger}_0)^{k-1}
(\hat{a}^\dagger_2)^{N-2k}
\hat{A}^{(2)\dagger}_2|{\rm vac}\rangle$ with $l=2$.
Since $F_z=F=2N_0-1$ is
forbidden, $F_z=2N-4k+3$ requires one more singlet pair to break up,
resulting in $l=5$ with the state 
$(\hat{A}^{(2)\dagger}_0)^{k-2}(\hat{a}^\dagger_2)^{N-2k-1}
\hat{A}^{(2)\dagger}_2\hat{A}^{(3)\dagger}_3|{\rm vac}\rangle$.
When $(pV^{\rm eff}+c_2/8)/c_1'-1/2$ falls between $2N-4k$ and
$2N-4(k-1)$, the lowest energy is the minimum of $E(F_z=2N-4k,l=0)$,
$E(F_z=2N-4k+1,l=3)$, $E(F_z=2N-4k+2,l=3)$, $E(F_z=2N-4k+3,l=5)$,
and $E(F_z=2N-4k+4,l=0)$. From Eq.~(\ref{e_fz_l}), we expect
that when $|c_2|/c_1'$ is large, nonzero $l$ pushes up
the energy significantly and cannot be the ground state, so that
$F_z$ increases stepwise with $\Delta F_z=4$.
When $|c_2|/c_1'$ is small, $F_z$ will increase with the step size
of $\Delta F_z=1$. This is confirmed by explicitly calculating
$E(F_z,l)$ using Eq.~(\ref{e_fz_l}),
and we obtain the diagrams in Fig.~\ref{fig:3}.

In the region $pV^{\rm eff}/c_1'<f_1(|c_2|/c_1')$ with
$f_1(x)=(x-20)(9x-4)(32x)^{-1}$ (see broken curves in Fig.~\ref{fig:3}),
    $F_z$ increases by taking every integer.
When $pV^{\rm eff}/c_1'>f_1(|c_2|/c_1')$, the values of $F_z=2N-4k+3$ are
suppressed. In the region $pV^{\rm eff}/c_1'>f_2(|c_2|/c_1')$ with
$f_2(x)=(x-20)(13x-20)(80x)^{-1}$, the values $F_z=2N-4k+1$ are
suppressed, and when $pV^{\rm eff}/c_1'>f_3(|c_2|/c_1')$ with
$f_3(x)=(x-20)(x-8)(8x)^{-1}$, the values $F_z=2N-4k+2$ are further
suppressed and the step size becomes 4.

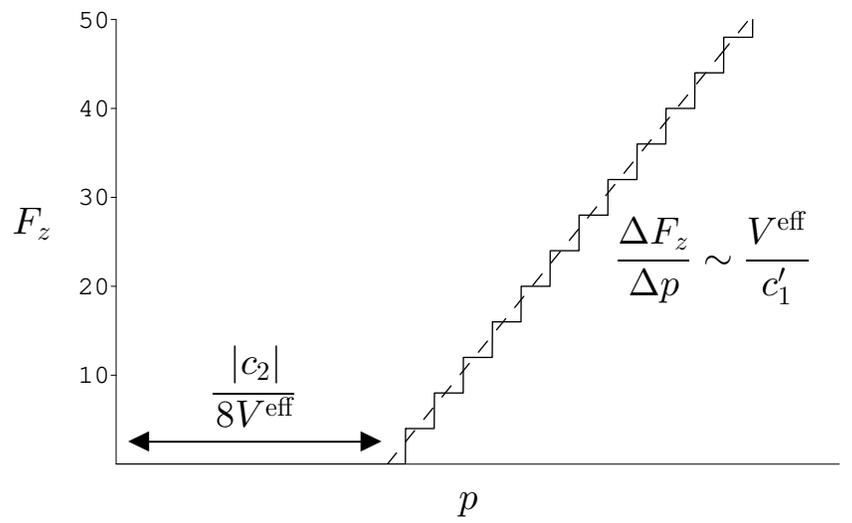
\begin{figure}
\caption{Typical dependence of the ground-state magnetization
on the applied magnetic-field strength, for $c_2<0$ and $c_1'>0$.
\label{fig:4}}
\end{figure}

While the averaged slope $\Delta F_z/\Delta p\sim V^{\rm eff}/c_1'$
coincides with that in MFT, the offset term $c_2/(8V^{\rm eff})$ in
Eq.~(\ref{e_fz_l}) (see also the broken lines in Fig.~\ref{fig:3}) 
makes a qualitative distinction from MFT, namely, the onset of the
magnetization displaces from $p=0$ to $p=|c_2|/(8V^{\rm eff})$. 
Note that the slope $ V^{\rm eff}/c_1'$ and the offset 
$|c_2|/(8V^{\rm eff})$ are determined by independent parameters. 
A typical behavior of the magnetic response 
when $|c_2|\gg c_1'$ is shown in Fig.~\ref{fig:4}.

\subsection{Property of ground states for $c_2<0$}

We now calculate the Zeeman-level populations of the
ground states for $c_2<0$. In MFT, the lowest-energy
states for $c_2<0$ have vanishing population in the $m=0,\pm 1$
levels. In contrast, the exact ground states derived in the preceding
subsection, $(\hat{A}^{(2)\dagger}_0)^{N_{\rm S}}
(\hat{a}^\dagger_2)^{n_{12}}
(\hat{A}^{(2)\dagger}_2)^{n_{22}}
(\hat{A}^{(3)\dagger}_3)^{n_{33}}|{\rm vac}\rangle$
with $n_{22}=0,1$ and $n_{33}=0,1$, 
have nonzero populations in the $m=0,\pm 1$
levels.  The exact forms for the averaged population
$\langle\hat{a}_m^\dagger\hat{a}_m\rangle$ are calculated as follows.
The above ground states have the form of $(\hat{A}_0^{(2)\dagger})^{N_{\rm
S}} |\phi\rangle\propto(\hat{\cal S}_+)^{N_{\rm S}} |\phi\rangle$ with
$|\phi\rangle$ being a state with a fixed number
   ($s\equiv n_{12}+2n_{22}+3n_{33}$) of bosons satisfying
$\hat{\cal S}_-|\phi\rangle=0$.
The average Zeeman population for the ground states,
\begin{equation}
\langle\hat{a}_m^\dagger\hat{a}_m\rangle
\equiv \frac{\langle \phi|(\hat{\cal S}_-)^{N_{\rm S}}
\hat{a}_m^\dagger\hat{a}_m
(\hat{\cal S}_+)^{N_{\rm S}} |\phi\rangle}
{\langle \phi|(\hat{\cal S}_-)^{N_{\rm S}}
(\hat{\cal S}_+)^{N_{\rm S}} |\phi\rangle},
\end{equation}
is then simply related to the average Zeeman populations for
the state $|\phi\rangle$ as
\begin{equation}
\langle\hat{a}_m^\dagger\hat{a}_m\rangle
=
\langle\hat{a}_m^\dagger\hat{a}_m\rangle_0
+\frac{N_{\rm S}}{s+5/2}
(\langle\hat{a}_m^\dagger\hat{a}_m\rangle_0+
\langle\hat{a}_{-m}^\dagger\hat{a}_{-m}\rangle_0+1),
\label{formula_zeeman1}
\end{equation}
where $\langle\hat{a}_m^\dagger\hat{a}_m\rangle_0\equiv
\langle \phi|\hat{a}_m^\dagger\hat{a}_m|\phi\rangle/\langle
\phi|\phi\rangle$.
The derivation of the formula (\ref{formula_zeeman1}) is given in
Appendix~\ref{app:zeeman}.
The formula implies that when $N_{\rm S}\gg s$,
the Zeeman populations of the ground states is 
sensitive to the form of $|\phi\rangle$.

With this formula, it is a straightforward task to calculate
average Zeeman-level populations for four types of
ground states,
$(\hat{A}^{(2)\dagger}_0)^{N_{\rm S}}
(\hat{a}^\dagger_2)^{n_{12}}
(\hat{A}^{(2)\dagger}_2)^{n_{22}}
(\hat{A}^{(3)\dagger}_3)^{n_{33}}|{\rm vac}\rangle$
with $n_{22}=0,1$ and $n_{33}=0,1$.
The exact result will be given in
Appendix~\ref{app:zeeman}.
A striking feature appears in the leading terms under
the condition $1\ll n_{12} \ll N_{\rm S}$.
The results are summarized as
\begin{equation}
\langle\hat{a}^\dagger_1\hat{a}_1\rangle
\sim
\langle\hat{a}^\dagger_{-1}\hat{a}_{-1}\rangle
\sim
N_{\rm S}(1+n_{33})/n_{12}
\end{equation}
and
\begin{equation}
\langle\hat{a}^\dagger_0\hat{a}_0\rangle
\sim
N_{\rm S}(1+2n_{22})/n_{12}.
\label{population_0}
\end{equation}
As seen in the preceding subsection, with the increase of magnetic field,
the ground state alternates among the four types of states. 
While this change causes a very small difference
in magnetization, it leads to large changes in the
average Zeeman-level populations, by a factor of 2 or 3.
The origin of this drastic change may be explained as follows.
Let us first consider the state
   $(\hat{a}^\dagger_2)^{n_{12}}|{\rm vac}\rangle$
with $n_{12}\ge 0$.
This state has no population in Zeeman levels $m=0,\pm1$.
When $\hat{A}^{(2)\dagger}_2$ is applied to this state,
the operator $\hat{a}_2^\dagger$ that appears in
$\hat{A}^{(2)\dagger}_2$ has effectively a large amplitude of
the order of $\sqrt{n_{21}}$. Hence the term
$\hat{a}_2^\dagger\hat{a}_0^\dagger$ is dominant, and
it approximately adds one boson to the $m=2$ level 
and one boson to the $m=0$ level.
Hence the $m=0$ population of the state
$\hat{A}^{(2)\dagger}_2(\hat{a}^\dagger_2)^{n_{12}}|{\rm vac}\rangle$
is close to unity. This change is then amplified by
a factor of $N_{\rm S}/n_{12}$ according to the
formula (\ref{formula_zeeman1}), leading to
Eq.~(\ref{population_0}).
Similarly, applying $\hat{A}^{(3)\dagger}_3$ effectively
results in adding of two bosons to the $m=2$ level and one boson 
to the $m=-1$ level through the dominant term
$(\hat{a}_2^\dagger)^2\hat{a}_{-1}^\dagger$.

\section{Low-lying excitation spectra}
\label{sec:Bogoliubov}

In this section we study the low-lying excitation spectrum of spin-2 BECs
in the thermodynamic limit using the Bogoliubov approximation. We shall  
see that the symmetry of each ground state discussed in Sec.~\ref{sec:MFT}
is reflected in the excitation spectrum.

\subsection{Effective Hamiltonian}

In the center-of-mass frame of the system BEC occurs in the ${\bbox{k}}={\bf
0}$
state. We therefore decompose the operators that appear in
Eq.~(\ref{spin2int2}) into the ${\bbox{k}}={\bf 0}$ components and the
${\bf
k}\neq{\bf 0}$ ones.
The first term on the right-hand side (rhs) of Eq.~(\ref{spin-1V1})
is rewritten as
\begin{eqnarray}
\sum_{\bbox{k}}:\hat{\rho}_{\bbox{k}}^\dagger\hat{\rho}_{\bbox{k}}:
=
:\hat{\rho}_{\bf 0}^\dagger\hat{\rho}_{\bf 0}:
+
\sum_{{\bbox{k}}\neq{\bf 0}}
:\hat{\rho}_{\bbox{k}}^\dagger\hat{\rho}_{\bbox{k}}:
=:\hat{N}^2:
+
\sum_{{\bbox{k}}\neq{\bf 0}}
:\hat{\rho}_{\bbox{k}}^\dagger\hat{\rho}_{\bbox{k}}:.
\label{decomp_rho}
\end{eqnarray}
If we ignore the terms that do not include the ${\bbox{k}}={\bf 0}$
components, we may approximate $\hat{\rho}_{{\bbox{k}}\neq{\bf 0}}$ as
\begin{eqnarray}
\hat{\rho}_{{\bbox{k}}\neq{\bf 0}}\simeq
\sum_m(\hat{a}_{{\bf 0},m}^\dagger\hat{a}_{{\bbox{k}},m}
+\hat{a}_{-{\bbox{k}},m}^\dagger\hat{a}_{{\bf 0},m})
=\hat{D}_{\bbox{k}}+\hat{D}_{-{\bbox{k}}}^\dagger,
\label{rhorho}
\end{eqnarray}
where
\begin{eqnarray}
\hat{D}_{\bbox{k}}\equiv\sum_m
\hat{a}_{{\bf 0},m}^\dagger\hat{a}_{{\bbox{k}},m}.
\label{D}
\end{eqnarray}
We thus obtain
\begin{eqnarray}
\sum_{\bbox{k}}:\hat{\rho}_{\bbox{k}}^\dagger\hat{\rho}_{\bbox{k}}:
\simeq
:\hat{N}^2:
+
\sum_{{\bbox{k}}\neq{\bf 0}}:(2\hat{D}_{\bbox{k}}^\dagger\hat{D}_{\bbox{k}}
+\hat{D}_{\bbox{k}}\hat{D}_{\bf -k}+\hat{D}_{\bbox{k}}^\dagger\hat{D}_{\bf
-k}^\dagger):.
\end{eqnarray}
Similarly, we may approximate the second term on the rhs of
Eq.~(\ref{spin-1V1}) as
\begin{eqnarray}
\sum_{\bbox{k}}:\hat{\bbox{F}}_{\bbox{k}}^\dagger\cdot
\hat{\bbox{F}}_{\bbox{k}}:
&\simeq&:\hat{\bbox{F}}^2:
+2\sum_{{\bbox{k}}\neq{\bf 0}}\sum_{mn}
\hat{\bbox{F}}\cdot{\bbox{f}}_{mn}
\hat{a}_{{\bbox{k}},m}^\dagger\hat{a}_{{\bbox{k}},n}
+\sum_{{\bbox{k}}\neq{\bf 0}}\sum_{ijmn}{\bbox{f}}_{ij}
\cdot{\bbox f}_{mn} \nonumber \\
& & \times
(2\hat{a}_{{\bf 0},i}^\dagger\hat{a}_{{\bf 0},n}
     \hat{a}_{{\bbox{k}},m}^\dagger\hat{a}_{{\bbox{k}},j}
+\hat{a}_{{\bf 0},i}^\dagger\hat{a}_{{\bf 0},m}^\dagger
     \hat{a}_{{\bbox{k}},j} \hat{a}_{-{\bbox{k}},n}
+\hat{a}_{{\bf 0},j}\hat{a}_{{\bf 0},n}
\hat{a}_{{\bbox{k}},i}^\dagger \hat{a}_{-{\bbox{k}},m}^\dagger),
\label{FF}
\end{eqnarray}
where
\begin{eqnarray}
\hat{\bbox{F}}=\sum_{mn}{\bbox{f}}_{mn}\hat{a}_{{\bf 0},m}^\dagger
\hat{a}_{{\bf 0},n}.
\label{F}
\end{eqnarray}
The third term on the rhs of Eq.~(\ref{spin-1V1}) is decomposed into
\begin{eqnarray}
\sum_{\bbox{k}}\hat{A}_{\bbox{k}}^\dagger\hat{A}_{\bbox{k}}
\simeq \frac{4}{5}\hat{S}_+\hat{S}_-
+\frac{2}{5}(\hat{S}_+\sum_{{\bbox{k}}\neq{\bf 0}}\sum_{m}(-1)^{m}
\hat{a}_{{\bbox{k}},m}\hat{a}_{{\bf -k},-m}+{\rm H.c.})
+
\frac{4}{5}\sum_{{\bbox{k}}\neq{\bf 0}}\sum_{mn}(-1)^{m+n}
\hat{a}_{{\bf 0},m}^\dagger\hat{a}_{{\bf 0},n}
     \hat{a}_{{\bbox{k}},-m}^\dagger\hat{a}_{{\bbox{k}},-n},
\label{SS}
\end{eqnarray}
where
\begin{eqnarray}
\hat{S}_-=\hat{S}_+^\dagger=\frac{1}{2}\sum_m(-1)^m\hat{a}_{{\bf 0},m}
\hat{a}_{{\bbox{k}},m},
\label{S}
\end{eqnarray}
and H.c. denotes Hermitian conjugates of the preceding terms.
Substituting Eqs.~(\ref{rhorho}), (\ref{FF}) and (\ref{SS})
into  Eq.~(\ref{spin-1V1}), we obtain
\begin{eqnarray}
\hat{V}&\simeq&\frac{c_0}{2V}:\hat{N}^2:+\frac{c_1}{2V}:\hat{\bbox{F}
}
^2:
+\frac{2c_2}{5V}\hat{S}_+\hat{S}_-
+\frac{c_0}{2V}\sum_{{\bbox{k}}\neq{\bf 0}}
:(2\hat{D}_{\bbox{k}}^\dagger\hat{D}_{{\bbox{k}}}
+\hat{D}_{\bbox{k}}\hat{D}_{-{\bbox{k}}}
+\hat{D}_{\bbox{k}}^\dagger\hat{D}_{\bf -k}^\dagger):
\nonumber \\
& &
+\frac{c_1}{V}\sum_{{\bbox{k}}\neq{\bf 0}}
\sum_{mn}\hat{\bbox{F}}\cdot
{\bbox{f}}_{mn}\hat{a}^\dagger_{{\bbox{k}},m}\hat{a}_{{\bbox{k}},n}
+\frac{c_1}{2V}
\sum_{{\bbox{k}}\neq{\bf 0}}\sum_{ijmn}{\bbox{f}}_{ij}
\cdot{\bbox{f}}_{mn}
(
2\hat{a}_{{\bf 0},i}^\dagger\hat{a}_{{\bf 0},n}
     \hat{a}_{{\bbox{k}},m}^\dagger\hat{a}_{{\bbox{k}},j}
+\hat{a}_{{\bf 0},i}^\dagger\hat{a}_{{\bf 0},m}^\dagger
     \hat{a}_{{\bbox{k}},j} \hat{a}_{-{\bbox{k}},n}
+\hat{a}_{{\bf 0},j}\hat{a}_{{\bf 0},n}
\hat{a}_{{\bbox{k}},i}^\dagger \hat{a}_{-{\bbox{k}},m}^\dagger)
\nonumber \\
& &
+\frac{c_2}{5V}\left(\hat{S}_+\sum_{{\bbox{k}}\neq{\bf 0}}\sum_{m}
(-1)^m\hat{a}_{{\bbox{k}},m}\hat{a}_{-{\bbox{k}},-m}
+{\rm H.c.} \right)
+\frac{2c_2}{5V}\sum_{{\bbox{k}}\neq{\bf 0}}\sum_{mn}(-1)^{m+n}
\hat{a}_{{\bf 0},m}^\dagger\hat{a}_{{\bf 0},n}
     \hat{a}_{{\bbox{k}},-m}^\dagger\hat{a}_{{\bbox{k}},-n}.
\label{spin2V}
\end{eqnarray}
In the Bogoliubov approximation we replace operators $\hat{a}_{{\bf 0},m}$
by c-numbers $\sqrt{N^{\rm BEC}}\zeta_m$, where $\zeta_m$'s denote the
complex mean-field amplitudes introduced in Sec.~\ref{sec:MFT} and 
$N^{\rm BEC}$ is the number of condensate bosons. 
Since $N^{\rm BEC}$ is smaller than the total number of bosons $N$ due to 
the interparticle interactions, we take into account the conservation of 
the total number of bosons through the relation
\begin{eqnarray}
N^{\rm BEC}+\sum_{{\bbox{k}}\neq{\bf 0}}\sum_{m}\hat{n}_{\bbox{k},m}=N.
\label{conservation_of_N}
\end{eqnarray}
Then Eq.~(\ref{H0}) becomes
\begin{eqnarray}
\hat{H}_0=-pN\langle\hat{f}_z\rangle+
\sum_{{\bbox{k}}\neq{\bf 0}}\sum_{m}
\left[\epsilon_{\bbox{k}}-p(m-\langle\hat{f}_z\rangle)\right]
\hat{n}_{\bbox{k},m},
\label{H00}
\end{eqnarray}
and Eq.~(\ref{spin2V}) becomes
\begin{eqnarray}
\hat{V}&\simeq& N\left(
\frac{c_0n}{2}+\frac{c_1n}{2}\langle \hat{\bbox{f}}\rangle^2
+\frac{2c_2n}{5}|\langle\hat{s}_-\rangle|^2\right)
-\left(c_1n\langle\hat{\bbox{f}}\rangle^2
+\frac{4c_2n}{5}|\langle\hat{s}_-\rangle|^2\right)\sum_{{\bbox{k}}\neq{\bf
0}}
\sum_{m}\hat{n}_{\bbox{k},m}
\nonumber \\
& &
+\frac{c_0n}{2}\sum_{{\bbox{k}}\neq{\bf 0}}\sum_{mn}
(2\zeta_m^*\zeta_n\hat{a}_{\bbox{k},n}^\dagger\hat{a}_{\bbox{k},m}
+\zeta_m^*\zeta_n^*\hat{a}_{\bbox{k},m}\hat{a}_{-\bbox{k},n}
+\zeta_m\zeta_n\hat{a}_{\bbox{k},m}^\dagger\hat{a}_{-\bbox{k},n}
)
\nonumber \\
& &
+c_1n\sum_{{\bbox{k}}\neq{\bf 0}}
\sum_{mn}\langle\hat{\bbox{f}}\rangle\cdot
{\bbox{f}}_{mn}\hat{a}^\dagger_{{\bbox{k}},m}\hat{a}_{{\bbox{k}},n}
+\frac{c_1n}{2}
\sum_{{\bbox{k}}\neq{\bf 0}}\sum_{ijmn}{\bbox{f}}_{ij}
\cdot{\bbox{f}}_{mn}
(
2\zeta_{i}^*\zeta_{n}\hat{a}_{{\bbox{k}},m}^\dagger\hat{a}_{{\bbox{k}},j}
+\zeta_{i}^*\zeta_{m}^*\hat{a}_{{\bbox{k}},j} \hat{a}_{-{\bbox{k}},n}
+\zeta_{j}\zeta_{n}
\hat{a}_{{\bbox{k}},i}^\dagger \hat{a}_{-{\bbox{k}},m}^\dagger)
\nonumber \\
& &
+\frac{c_2n}{5}\left(\langle\hat{s}_-\rangle^*
\sum_{{\bbox{k}}\neq{\bf 0}}\sum_{m}
(-1)^m\hat{a}_{{\bbox{k}},m}\hat{a}_{-{\bbox{k}},-m}
+{\rm H.c.} \right)
+\frac{2c_2n}{5}\sum_{{\bbox{k}}\neq{\bf 0}}\sum_{mn}(-1)^{m+n}
\zeta_{m}^*\zeta_{n}
\hat{a}_{\bbox{k},-m}^\dagger\hat{a}_{{\bbox{k}},-n},
\label{spin2VV}
\end{eqnarray}
where $n\equiv N/V$, and the definitions 
of $\langle\hat{f}_z\rangle$, $\langle\hat{\bbox{f}}\rangle$,
and $\langle\hat{s}_-\rangle$ are given in Eqs.~(\ref{f})--(\ref{s-}).
Equations~(\ref{H00}) and (\ref{spin2VV}) constitute our basic Hamiltonian
in the following discussions.
We use this Hamiltonian to examine low-lying excitation spectra
for each phase.

\subsection{Excitation spectrum of a ferromagnetic BEC}
\label{sec:FMB}

Let us first examine the excitation spectrum of a ferromagnetic phase
in which BEC occurs in the $m=2$ state. Then $\zeta_m=2\delta_{m,2}$
and Eqs.~(\ref{H00}) and (\ref{spin2VV}) become
\begin{eqnarray}
\hat{H}_0
= -2p\hat{N}
+\sum_{{\bbox{k}}\neq{\bf 0},m}
\left[\epsilon_{\bbox{k}}+(2-m)p\right]\hat{n}_{{\bbox{k}},m},
\label{H0f}
\end{eqnarray}
and
\begin{eqnarray}
\hat{V}&\simeq&\frac{1}{2}g_4n\hat{N}
+n\sum_{{\bbox{k}}\neq{\bf 0}}\left[
c_0\hat{n}_{{\bbox{k}},2}+2c_1
(2\hat{n}_{{\bbox{k}},2}-2\hat{n}_{{\bbox{k}},0}-3\hat{n}_{{\bbox{k}},-1}
-4\hat{n}_{{\bbox{k}},-2})
+\frac{2}{5}c_2\hat{n}_{{\bbox{k}},-2}
+\frac{g_4}{2}(\hat{a}_{{\bbox{k}},2}\hat{a}_{-{\bbox{k}},2}+
\hat{a}_{{\bbox{k}},2}^\dagger\hat{a}_{-{\bbox{k}},2}^\dagger)
\right],
\end{eqnarray}
respectively, where $g_4 = c_0+4c_1$.
The total Hamiltonian is therefore given by
\begin{eqnarray}
\hat{H}^{\rm F}&=& \frac{1}{2}g_4n\hat{N}-2p\hat{N}
\nonumber \\
& & +\sum_{{\bbox{k}}\neq{\bf 0}}
\left[\left(\epsilon_{\bbox{k}}
+g_4n\right)\hat{n}_{{\bbox{k}},2}
+\frac{1}{2}g_4n(\hat{a}_{{\bbox{k}},2}\hat{a}_{-{\bbox{k}},2}+
\hat{a}_{{\bbox{k}},2}^\dagger\hat{a}_{-{\bbox{k}},2}^\dagger) \right]
\nonumber \\
& & +\sum_{{\bbox{k}}\neq{\bf 0}}
\left[(\epsilon_{\bbox{k}}+p)\hat{n}_{{\bbox{k}},1}
+\left(\epsilon_{\bbox{k}}+2p-4c_1n\right)\hat{n}_{{\bbox{k}},0}
\right.
\nonumber \\
& & \left.
+\left(\epsilon_{\bbox{k}}+3p-6c_1n\right)\hat{n}_{{\bbox{k}},-1}
+\left(\epsilon_{\bbox{k}}+4p-8c_1n+2c_2n/5 \right)\hat{n}_{{\bbox{k}},-2}
\right].
\label{HFM}
\end{eqnarray}
The second line gives the Bogoliubov spectrum
\begin{eqnarray}
E^{\rm F}_{m=2,\bbox{k}}=\sqrt{\epsilon_{\bbox{k}}^2+2g_4n\epsilon_{\bf
k}},
\label{FBog}
\end{eqnarray}
while other terms give single-particle spectra:
\begin{eqnarray}
E^{\rm F}_{m=1,\bbox{k}}&=& \epsilon_{\bbox{k}}+p  \label{FS1}\\
E^{\rm F}_{m=0,\bbox{k}}&=&\epsilon_{\bbox{k}}+2p-4c_1n \label{FS0}\\
E^{\rm F}_{m=-1,\bbox{k}}&=&\epsilon_{\bbox{k}}+3p-6c_1n \label{FSM1}\\
E^{\rm F}_{m=-2,\bbox{k}}&=&\epsilon_{\bbox{k}}+4p-8c_1n+2c_2n/5
\label{FSM2}
\end{eqnarray}
For the Bogoliubov excitation energy to be positive, we must have
\begin{eqnarray}
g_4=\frac{4\pi\hbar^2}{M}a_4>0.
\end{eqnarray}
That is, the s-wave scattering length for the total spin-4 channel
must be positive. This condition is the same as that required for
the ferromagnetic mean field to be stable, that is, the first term
on the rhs of Eq.~(\ref{HFM}) being positive.
For the single-particle excitation energies to be positive, we must
have $p>2c_1n$ and $p>(2c_1-c_2/10)n$.
These conditions are the same as those in~(\ref{condF}) for which
the ferromagnetic phase is the lowest-energy mean field
(note that $\tilde{p}\simeq 2p/n$).

We note that the Bogoliubov spectrum (\ref{FBog}) is independent of applied
magnetic field and remains massless in its presence. This Goldstone mode
is a consequence of the global U(1) gauge invariance due to the conservation
of the total number of bosons, as discussed in Sec.~\ref{sec:MFFM}.

\subsection{Excitation spectrum of an antiferromagnetic BEC}
\label{sec:AFB}

Let us next examine the excitation spectrum of an antiferromagnetic phase
in which the order parameter is given by (\ref{AFspinor}).
Making the replacements
\begin{eqnarray}
\zeta_{2}=
e^{i\phi_{2}}
\sqrt{\frac{1}{2}+\frac{\langle\hat{f}_z\rangle}{4}}, \ \
\zeta_{-2}=
e^{i\phi_{-2}}
\sqrt{\frac{1}{2}-\frac{\langle\hat{f}_z\rangle}{4}}, \ \
\zeta_{0}=\zeta_{\pm1}=0,
\label{fsd}
\end{eqnarray}
and substituting Eq.~(\ref{fsd}) into Eq.~(\ref{spin2VV}), together with
Eq.~(\ref{H00}), we obtain the total Hamiltonian of an antiferromagnetic BEC:
\begin{eqnarray}
\hat{H}^{\rm AF}&=&\frac{1}{2}(c_0+c_2/5)nN
+\frac{1}{2}(c_1-c_2/20)nN\langle\hat{f}_z\rangle^2
-pN\langle\hat{f}_z\rangle
\nonumber \\
& &
+\sum_{{\bbox{k}}\neq{\bf 0}}\left\{
\left[\epsilon_{\bbox{k}}+p(\langle\hat{f}_z\rangle-2)+c_0n|\zeta_{2}|^2
+c_1n(4|\zeta_{2}|^2+2\langle\hat{f}_z\rangle-\langle\hat{f}_z\rangle^2)
+\frac{2c_2n}{5}(|\zeta_{-2}|^2-2|\zeta_{2} \zeta_{-2}|^2)
\right]\hat{n}_{{\bbox{k}},2}
\right. \nonumber \\
& & \ \ \
+\left[\epsilon_{\bbox{k}}+p(\langle\hat{f}_z\rangle+2)+c_0n|\zeta_{-2}|^2
+c_1n(4|\zeta_{-2}|^2-2\langle\hat{f}_z\rangle-\langle\hat{f}_z\rangle^2)
+\frac{2c_2n}{5}(|\zeta_{2}|^2-2|\zeta_{2} \zeta_{-2}|^2)
\right]\hat{n}_{{\bbox{k}},-2}
\nonumber \\
& & \left. \ \ \
+\left[
\frac{1}{2}g_4n(\zeta_{2}^{*2}\hat{a}_{{\bbox{k}},2}\hat{a}_{-{\bbox{k}},2}
+\zeta_{-2}^{*2}\hat{a}_{{\bbox{k}},-2}\hat{a}_{-{\bbox{k}},-2})
+
(c_0-4c_1+2c_2/5)n(
\zeta_{2}^*\zeta_{-2}^*\hat{a}_{{\bbox{k}},2}\hat{a}_{-{\bbox{k}},-2}+
\zeta_{2}
\zeta_{-2}^*\hat{a}_{{\bbox{k}},2}^\dagger\hat{a}_{{\bbox{k}},-2}) +{\rm
H.c.}\right]
\right\}
\nonumber \\
& &
+\sum_{{\bbox{k}}\neq{\bf 0}}\left\{
\left[\epsilon_{\bbox{k}}+p(\langle\hat{f}_z\rangle-1)
+c_1n(2|\zeta_{2}|^2+\langle\hat{f}_z\rangle-\langle\hat{f}_z\rangle^2)
-\frac{4c_2n}{5}|\zeta_{2} \zeta_{-2}|^2
\right]\hat{n}_{{\bbox{k}},1}
\right. \nonumber \\
& & \ \ \
+\left[\epsilon_{\bbox{k}}+p(\langle\hat{f}_z\rangle+1)
+c_1n(2|\zeta_{-2}|^2-\langle\hat{f}_z\rangle-\langle\hat{f}_z\rangle^2)
-\frac{4c_2n}{5}|\zeta_{2} \zeta_{-2}|^2
\right]\hat{n}_{{\bbox{k}},-1}
\nonumber \\
& & \left. \ \ \
+2(c_1-c_2/5)n(\zeta_{2}^*\zeta_{-2}^*\hat{a}_{{\bbox{k}},1}\hat{a}_{-{\bbox
{k}},-1}
+{\rm H.c.})
\right\}
\nonumber \\
& &
+\sum_{{\bbox{k}}\neq{\bf 0}}\left\{
\left[\epsilon_{\bf
k}+p\langle\hat{f}_z\rangle-c_1n\langle\hat{f}_z\rangle^2-
\frac{4c_2n}{5}|\zeta_{2} \zeta_{-2}|^2\right]
\hat{n}_{{\bbox{k}},0}
+\frac{c_2n}{5}(\zeta_{2}^*\zeta_{-2}^*\hat{a}_{{\bbox{k}},0}
\hat{a}_{-{\bbox{k}},0}+{\rm
H.c.})
\right\}.
\label{HAFM1}
\end{eqnarray}
This Hamiltonian may be simplified using the relation
$p\simeq\langle\hat{f}_z\rangle(c_1-c_2/20)n$, giving
\begin{eqnarray}
\hat{H}^{\rm AF}&=&\frac{1}{2}(c_0+c_2/5)nN
-\frac{1}{2}pN\langle\hat{f}_z\rangle
\nonumber \\
& &
+\sum_{{\bbox{k}}\neq{\bf 0}}\left\{
(\epsilon_{\bbox{k}}+g_4n|\zeta_{2}|^2)\hat{n}_{{\bbox{k}},2}
+(\epsilon_{\bbox{k}}+g_4n|\zeta_{-2}|^2)\hat{n}_{{\bbox{k}},-2}
\right. \nonumber \\
\nonumber \\
& & \left. \ \ \
+\left[
\frac{g_4n}{2}(\zeta_{2}^{*2}\hat{a}_{{\bbox{k}},2}\hat{a}_{-{\bbox{k}},2}
+\zeta_{-2}^{*2}\hat{a}_{{\bbox{k}},-2}\hat{a}_{-{\bbox{k}},-2})
+
(c_0-4c_1+2c_2/5)n
(\zeta_{2}^*\zeta_{-2}^*\hat{a}_{{\bbox{k}},2}\hat{a}_{-{\bbox{k}},-2}+
\zeta_{2}
\zeta_{-2}^*\hat{a}_{{\bbox{k}},2}^\dagger\hat{a}_{{\bbox{k}},-2}) +{\rm
H.c.}\right]
\right\}
\nonumber \\
& &
+\sum_{{\bbox{k}}\neq{\bf 0}}\left\{
\left[\epsilon_{\bbox{k}}
+(c_1-c_2/5)n+\frac{1}{2}(c_1+c_2/10)n\langle\hat{f}_z\rangle
\right]\hat{n}_{{\bbox{k}},1}
+
\left[\epsilon_{\bbox{k}}
+(c_1-c_2/5)n-\frac{1}{2}(c_1+c_2/10)n\langle\hat{f}_z\rangle
\right]\hat{n}_{{\bbox{k}},-1} \right.
\nonumber \\
& & \left. \ \ \
+2(c_1-c_2/5)n(\zeta_{2}^*\zeta_{-2}^*\hat{a}_{{\bbox{k}},1}\hat{a}_{-{\bbox
{k}},-1}
+{\rm H.c.})
\right\}
\nonumber \\
& &
+\sum_{{\bbox{k}}\neq{\bf 0}}\left\{
(\epsilon_{\bbox{k}}-c_2n/5)
\hat{n}_{{\bbox{k}},0}
+\frac{c_2n}{5}(\zeta_{2}^*\zeta_{-2}^*\hat{a}_{{\bbox{k}},0}
\hat{a}_{-{\bbox{k}},0}+{\rm
H.c.})
\right\}.
\label{HAFM}
\end{eqnarray}
This result shows that the eigenmodes are classified into three categories:
the $m=0$ mode, the coupled $m=\pm1$ modes, and the coupled $m=\pm2$ modes.
Below we analyze each of them.

\subsubsection{The spin-0 quasiparticles}
\label{sec:spin-0}

The Hamiltonian (\ref{HAFM}) shows that the $m=0$ mode is decoupled from
other modes even in the presence of magnetic field.
This part of the Hamiltonian can readily be diagonalized to give
\begin{eqnarray}
E_{{\bbox{k}},0}=
\sqrt{\epsilon_{\bbox{k}}^2+\frac{2|c_2|n}{5}\epsilon_{\bbox{k}}
+\left(\frac{c_2n}{10}\right)^2\langle\hat{f}_z\rangle^2}.
\label{AF0}
\end{eqnarray}
We note that the spectrum~(\ref{AF0}) becomes massive in the presence
of magnetic field.

\subsubsection{The spin-1 quasiparticles}

The $m=1$ and $m=-1$ modes are coupled in the Hamiltonian (\ref{HAFM}),
and the eigenenergies can be obtained by diagonalizing the following 
Hamiltonian:
\begin{eqnarray}
\hat{h}_1^{\rm AF}=\sum_{{\bbox{k}}\neq{\bf 0}}\left\{
\epsilon_{{\bbox{k}},1}\hat{n}_{{\bbox{k}},1}+
\epsilon_{{\bbox{k}},-1}\hat{n}_{-{\bbox{k}},-1}
+(\delta\hat{a}_{{\bbox{k}},1}\hat{a}_{-{\bbox{k}},-1}+{\rm H.c.})
\right\},
\label{h1}
\end{eqnarray}
where $\delta\equiv 2(c_1-c_2/5)n\zeta_{2}^*\zeta_{-2}^*$ and
\begin{eqnarray}
\epsilon_{{\bbox{k}},\pm1}\equiv\epsilon_{\bbox{k}}
+(c_1-c_2/5)n\pm\frac{1}{2}(c_1+c_2/10)n\langle\hat{f}_z\rangle
\end{eqnarray}
The dispersion relation can be found by writing down the equations of
motion for $\hat{a}_{{\bbox{k}},1}$ and $\hat{a}_{-{\bbox{k}},-1}^\dagger$
and seeking for the solution of the form
$\exp({\mp iE_{{\bbox{k}},\pm1}t/\hbar})$, with
the
result
\begin{eqnarray}
E_{{\bbox{k}},\pm1}&=&\pm\frac{1}{2}\left(c_1+\frac{c_2}{10}\right)n
\langle\hat{f}_z\rangle
+\sqrt{\epsilon_{\bbox{k}}^2+2\left(c_1-\frac{c_2}{5}\right)n\epsilon_{\bf
k}
+\frac{1}{4}\left(c_1-\frac{c_2}{5}\right)^2n^2\langle\hat{f}_z\rangle^2}.
\label{spin-1}
\end{eqnarray}
The excitation spectra (\ref{spin-1}) become massive in the presence of
magnetic field.
The positivity of this energy is guaranteed by the conditions
$c_1-c_2/20>p/2n>0$ and $c_2<0$ which are required for the
antiferromagnetic phase to be the lowest-lying mean field 
(see condition (\ref{condAF})).

\subsubsection{The spin-2 quasiparticles}

The $m=2$ and $m=-2$ modes are coupled in the Hamiltonian (\ref{HAFM}),
and the relevant part of the Hamiltonian reads
\begin{eqnarray}
\hat{h}_2&=&\sum_{{\bbox{k}}\neq{\bf 0}}
\left\{
\epsilon_{{\bbox{k}},2}\hat{n}_{{\bbox{k}},2}+
\epsilon_{{\bbox{k}},-2}\hat{n}_{{\bbox{k}},-2}
+\frac{g_4n}{2}
(\zeta_{2}^{*2}\hat{a}_{{\bbox{k}},2} \hat{a}_{-{\bbox{k}},2}
+\zeta_{-2}^{*2}\hat{a}_{{\bbox{k}},-2}\hat{a}_{-{\bbox{k}},-2}+{\rm
H.c.})
\right.
\nonumber \\
& &\left.
+(c_0-4c_1+2c_2/5)n
(\zeta_{2}^*\zeta_{-2}^* \hat{a}_{{\bbox{k}},2}\hat{a}_{-{\bbox{k}},-2}+
\zeta_{2} \zeta_{-2}^*
\hat{a}_{{\bbox{k}},2}^\dagger\hat{a}_{{\bbox{k}},-2} +{\rm
H.c.})\right\},
\label{h2af}
\end{eqnarray}
where
\begin{eqnarray}
\epsilon_{{\bbox{k}},2} =\epsilon_{\bbox{k}}+g_4n|\zeta_{2}|^2, \ \ \
\epsilon_{{\bbox{k}},-2}=\epsilon_{\bbox{k}}+g_4n|\zeta_{-2}|^2.
\end{eqnarray}
By unitary transformations
$\hat{a}_{\pm{\bbox{k}},2}\rightarrow \hat{a}_{\pm{\bbox{k}},2}e^{i\phi_2}$ 
and
$\hat{a}_{\pm{\bbox{k}},-2}\rightarrow\hat{a}_{\pm{\bbox{k}},-2}
e^{i\phi_{-2}}$,
Eq.~(\ref{h2af}) reduces to
\begin{eqnarray}
\hat{h}_2^{\rm AF}&=&\sum_{{\bbox{k}}\neq{\bf 0}}
\left\{
\epsilon_{{\bbox{k}},2}\hat{n}_{{\bbox{k}},2}+
\epsilon_{{\bbox{k}},-2}\hat{n}_{{\bbox{k}},-2}
+\frac{1}{2}
(\alpha\hat{a}_{{\bbox{k}},2} \hat{a}_{-{\bbox{k}},2}
+\beta \hat{a}_{{\bbox{k}},-2}\hat{a}_{-{\bbox{k}},-2}+{\rm H.c.})
+\gamma
(\hat{a}_{{\bbox{k}},2}\hat{a}_{-{\bbox{k}},-2}+
\hat{a}_{{\bbox{k}},2}^\dagger\hat{a}_{{\bbox{k}},-2}
+{\rm H.c.})\right\},
\label{h2af2}
\end{eqnarray}
where $\alpha\equiv g_4n|\zeta_{2}|^2$, $\beta\equiv g_4n|\zeta_{-2}|^2$,
and
$\gamma\equiv(c_0-4c_1+2c_2/5)n|\zeta_{2} \zeta_{-2}|$.
This Hamiltonian can be diagonalized by writing down the Heisenberg
equations of motion for
$\hat{a}_{{\bbox{k}},2}+\hat{a}_{-{\bbox{k}},2}^\dagger$
and $\hat{a}_{{\bbox{k}},-2}+\hat{a}_{-{\bbox{k}},-2}^\dagger$ as
\begin{eqnarray}
& &
(i\hbar)^2\frac{d^2}{dt^2}(\hat{a}_{{\bbox{k}},2}
+\hat{a}_{-{\bbox{k}},2}^\dagger)
=
\epsilon_{\bbox{k}}(\epsilon_{\bbox{k}}+2\alpha)
(\hat{a}_{{\bbox{k}},2}+\hat{a}_{-{\bbox{k}},2}^\dagger)
+2\gamma\epsilon_{\bbox{k}}(\hat{a}_{{\bbox{k}},-2}+\hat{a}_{-{\bbox{k}},-2}
^\dagger),
\nonumber \\
& &
(i\hbar)^2\frac{d^2}{dt^2}(\hat{a}_{{\bbox{k}},-2}+\hat{a}_{-{\bf
k},-2}^\dagger)
=\epsilon_{\bbox{k}}(\epsilon_{\bbox{k}}+\beta)
(\hat{a}_{{\bbox{k}},-2}+\hat{a}_{-{\bbox{k}},-2}^\dagger)
+2\gamma\epsilon_{\bbox{k}}
(\hat{a}_{{\bbox{k}},2}+\hat{a}_{-{\bbox{k}},2}^\dagger).
\end{eqnarray}
By assuming that
$\hat{a}_{{\bbox{k}},\pm2}+\hat{a}_{-{\bbox{k}},\pm2}^\dagger
\propto \exp({-{i}E^{\rm AF}_{{\bbox{k}},\pm2} t/{\hbar}})$,
we obtain the following dispersion relations:
\begin{eqnarray}
(E^{\rm AF}_{{\bbox{k}},\pm2})^2=
\epsilon_{\bbox{k}}\left[
\epsilon_{\bbox{k}}+g_4n\pm
g_4n\sqrt{\frac{\langle\hat{f}_z\rangle^2}{4}+\left[1-\frac{8}{g_4}
\left(c_1-\frac{c_2}{20}\right)\right]^2\left(1-\frac{\langle\hat{f}_z
\rangle^2}{4}\right)}
\right].
\label{dis2}
\end{eqnarray}
The positivity of this energy is met if the conditions $c_1-c_2/20>0$
and $c_0+c_2/5>0$ are satisfied.
The former condition is met whenever the antiferromagnetic phase is the
lowest-energy state (see (\ref{condAF})), while
the latter condition is required for the antiferromagnetic phase to be
mechanically stable, that is,
the first term on the rhs of Eq.~(\ref{HAFM}) is positive.
We note that the dispersion relations (\ref{dis2}) are massless even in the
presence of the magnetic field.
They are the Goldstone modes associated with the U(1) gauge symmetry
and the relative gauge symmetry (the rotational symmetry about the
direction of the applied magnetic field)
that are manifest in the mean-field
solution discussed in Sec.~\ref{sec:MFAF}.

At zero magnetic field,
Eqs.~(\ref{AF0}), (\ref{spin-1}) and (\ref{dis2}) reduces to
\begin{eqnarray}
(E^{\rm AF}_{{\bbox{k}},0})^2&=&
\epsilon_{\bbox{k}}\left[\epsilon_{\bbox{k}}-(2c_2/5)n\right],
\\
(E^{\rm AF}_{{\bbox{k}},\pm 1})^2&=&
\epsilon_{\bbox{k}}\left[\epsilon_{\bbox{k}}+2(c_1-c_2/5)n\right],
\\
(E^{\rm AF}_{{\bbox{k}},\pm2})^2&=&
\cases{
\epsilon_{\bbox{k}}\left[\epsilon_{\bbox{k}}+2(c_0+c_2/5)n\right] ,  \cr
\epsilon_{\bbox{k}}\left[\epsilon_{\bbox{k}}+8(c_1-c_2/20)n\right],  \cr}
\end{eqnarray}
implying that all the five excitations are Goldstone modes.
This reflects the fact that in the absence of the magnetic field the
ground state is degenerate with respect to five continuous variables
(see Eq.~(\ref{AFspinor2})).

\subsection{Excitation spectrum of a cyclic BEC}
\label{sec:CB}

We consider the case of Eq.~(\ref{cyclic_mfs}), namely,
\begin{eqnarray}
\zeta_{\pm2}=\frac{1}{2}\left(1\pm a\right)e^{i\phi_{\pm2}}, \ \
\zeta_{\pm1}=0, \ \
\zeta_0=\sqrt{\frac{1-a^2}{2}}e^{i\phi_0},
\label{cyclic_mfs2}
\end{eqnarray}
where $a\equiv\langle\hat{f}_z\rangle/2$ and
$\phi_2+\phi_{-2}-2\phi_0=\pm\pi$.
Because $\langle\hat{s}_-\rangle=0$ and $\langle\hat{f}_\pm\rangle=0$,
the interaction Hamiltonian (\ref{spin2VV}) reduces to
\begin{eqnarray}
\hat{V}&\simeq&\frac{c_0n}{2}+\frac{1}{2}pN\langle\hat{f}_z\rangle
-c_1n\langle\hat{f}_z\rangle^2\sum_{{\bbox{k}}\neq{\bf 0}}\sum_m
\hat{n}_{{\bbox{k}},m}
+\frac{c_0}{2V}\sum_{{\bbox{k}}\neq{\bf 0}}:
(\hat{D}_{\bbox{k}}\hat{D}_{-{\bbox{k}}}+\hat{D}_{\bbox{k}}^\dagger\hat{D}_
{\bbox{k}}
+{\rm H.c.}): + c_1n\langle\hat{f}_z\rangle\sum_{{\bbox{k}}\neq{\bf 0}}
\sum_mm\hat{n}_{{\bbox{k}},m}
\nonumber \\
& &
+\frac{c_1}{2V}
\sum_{{\bbox{k}}\neq{\bf 0}}\sum_{ijmn}{\bbox{f}}_{ij}{\bbox{f}}_{mn}
(
\hat{a}_{{\bf 0},i}^\dagger\hat{a}_{{\bf 0},n}
     \hat{a}_{{\bbox{k}},m}^\dagger\hat{a}_{{\bbox{k}},j}
+\hat{a}_{{\bf 0},i}^\dagger\hat{a}_{{\bf 0},m}^\dagger
     \hat{a}_{{\bbox{k}},j} \hat{a}_{-{\bbox{k}},n}+{\rm H.c.})
\nonumber \\
& &
+\frac{2c_2}{5V}\sum_{{\bbox{k}}\neq{\bf 0}}\sum_{mn}(-1)^{m+n}
\hat{a}_{{\bf 0},m}^\dagger\hat{a}_{{\bf 0},n}
     \hat{a}_{{\bbox{k}},-m}^\dagger\hat{a}_{{\bbox{k}},-n}.
\label{Vcyclic}
\end{eqnarray}
Substituting Eq.~(\ref{cyclic_mfs2}) into this, performing unitary
transformations
$\hat{a}_{\pm{\bbox{k}},m}\rightarrow \hat{a}_{\pm{\bbox{k}},m}e^{i\phi_m}$,
where $\zeta_m=|\zeta_m|e^{i\phi_m}$ and $\phi_2+\phi_{-2}-2\phi_0=\pi$,
and combining the result with Eq.~(\ref{H00}), we obtain
\begin{eqnarray}
\hat{H}^{\rm C}&=&\frac{1}{2}c_0nN-\frac{1}{2}pN\langle\hat{f}_z\rangle
+\sum_{{\bbox{k}}\neq{\bf 0},m} A_{{\bbox{k}},m}\hat{n}_{{\bbox{k}},m}
\nonumber \\
& & +\sum_{{\bbox{k}}\neq{\bf 0}}
\left\{
\frac{1}{2}(\alpha+\beta)(|\zeta_2|^2\hat{a}_{{\bbox{k}},2}\hat{a}_{-{\bbox
{k}},2}
+|\zeta_{-2}|^2\hat{a}_{{\bbox{k}},-2}\hat{a}_{-{\bbox{k}},-2})
+\frac{\alpha}{2}|\zeta_0|^2\hat{a}_{{\bbox{k}},0}\hat{a}_{-{\bbox{k}},0}
+(\alpha-\beta)|\zeta_2\zeta_{-2}|\hat{a}_{{\bbox{k}},2}\hat{a}_{-{\bbox{k}}
,
-2}
\right. \nonumber \\
& &
+(\alpha-\beta+\gamma)|\zeta_2\zeta_{-2}|
\hat{a}_{{\bbox{k}},2}^\dagger\hat{a}_{{\bbox{k}},-2}
+\alpha|\zeta_0|(|\zeta_2|\hat{a}_{{\bbox{k}},2}\hat{a}_{-{\bbox{k}},0}
+|\zeta_{-2}|\hat{a}_{{\bbox{k}},0}\hat{a}_{-{\bbox{k}},-2})
\nonumber \\
& & \left.
+|\zeta_0|(\alpha|\zeta_2|-\gamma|\zeta_{-2}|)
\hat{a}_{{\bbox{k}},2}^\dagger\hat{a}_{{\bbox{k}},0}
+|\zeta_0|(\alpha|\zeta_{-2}|-\gamma|\zeta_2|)
\hat{a}_{{\bbox{k}},0}^\dagger\hat{a}_{{\bbox{k}},-2}
+{\rm H.c.}
\right\}
\nonumber \\
& & +\frac{\beta}{4}\sum_{{\bbox{k}}\neq{\bf 0}}\left\{
2\zeta_0^{*2}\hat{a}_{{\bbox{k}},1}\hat{a}_{-{\bbox{k}},-1}
+\sqrt{6}\zeta_0^*
(\zeta_2^*\hat{a}_{{\bbox{k}},1}\hat{a}_{-{\bbox{k}},1}
+\zeta_{-2}^*\hat{a}_{{\bbox{k}},-1}\hat{a}_{-{\bbox{k}},-1})
+\sqrt{6}(\zeta_2^*\zeta_0+\zeta_0^*\zeta_{-2})
\hat{a}_{{\bbox{k}},-1}^\dagger
\hat{a}_{{\bbox{k}},1}+{\rm H.c.}
\right\},
\label{Hcyclic}
\end{eqnarray}
where $\alpha\equiv c_0n$, $\beta\equiv 4c_1n$, $\gamma\equiv 2c_2n/5$, and
\begin{eqnarray}
A_{{\bbox{k}},\pm 2}&=&\epsilon_{\bbox{k}}
+(\alpha+\beta)|\zeta_{\pm2}|^2+\gamma|\zeta_{\mp2}|^2,
\nonumber \\
A_{{\bbox{k}},\pm 1}&=&\epsilon_{\bbox{k}}
+\frac{\beta}{4}(2|\zeta_{\pm2}|^2+3|\zeta_0|^2),
\nonumber \\
A_{{\bbox{k}},0}&=&\epsilon_{\bbox{k}}
+(\alpha+\gamma)|\zeta_0|^2.
\end{eqnarray}
It can be seen from the Hamiltonian (\ref{Hcyclic}) that
there are two separate sets of coupled modes, that is, the $m=\pm1$ modes
and the $m=0,\pm2$ modes.

\subsubsection{The $m=\pm1$ coupled modes}

The equations of motion governing the $m=\pm1$ coupled modes are given by
\begin{eqnarray}
i\hbar\frac{d}{dt}\hat{a}_{{\bbox{k}},1}&=&
A_{{\bbox{k}},1}\hat{a}_{{\bbox{k}},1}+D_2\hat{a}_{-{\bbox{k}},1}^\dagger
+B\hat{a}_{-{\bbox{k}},-1}^\dagger
+C\hat{a}_{{\bbox{k}},-1} \nonumber \\
i\hbar\frac{d}{dt}\hat{a}_{-{\bbox{k}},1}^\dagger &=&
-A_{{\bbox{k}},1}\hat{a}_{-{\bbox{k}},1}\dagger-D_2^*\hat{a}_{{\bbox{k}},1}
-B^*\hat{a}_{{\bbox{k}},-1}
-C^*\hat{a}_{-{\bbox{k}},-1}^\dagger \nonumber \\
i\hbar\frac{d}{dt}\hat{a}_{{\bbox{k}},-1}&=&
A_{{\bbox{k}},-1}\hat{a}_{{\bbox{k}},-1}+D_{-2}
\hat{a}_{-{\bbox{k}},-1}^\dagger
+B\hat{a}_{-{\bbox{k}},1}^\dagger
+C^*\hat{a}_{{\bbox{k}},1} \nonumber \\
i\hbar\frac{d}{dt}\hat{a}_{-{\bbox{k}},-1}^\dagger&=&
-A_{{\bbox{k}},-1}\hat{a}_{-{\bbox{k}},-1}^\dagger
-D_{-2}^*\hat{a}_{{\bbox{k}},-1}
-B^*\hat{a}_{{\bbox{k}},1}
-C\hat{a}_{-{\bbox{k}},1}^\dagger,
\end{eqnarray}
where
$B\equiv\beta\zeta_0^2/2$,
$C\equiv\sqrt{6}\beta(\zeta_2\zeta_0^*+\zeta_0\zeta_{-2}^*)/4$,
$D_{\pm 2}\equiv\sqrt{6}\beta\zeta_0\zeta_{\pm2}/2$.
The eigenenergies of these modes are given by
\begin{eqnarray}
(E_{\pm1}^{\rm C})^2&=&\epsilon_{\bf
k}^2+\beta\left(1-\frac{\langle\hat{f}_z\rangle^2}{8}\right)\epsilon_{\bbox{
k}}
+\frac{\beta^2}{32}\langle\hat{f}_z\rangle^2
\pm\frac{\beta\langle\hat{f}_z\rangle}{2}
\left\{
\left(\epsilon_{\bbox{k}}+\frac{\beta}{4}\right)
\left[\left(1-\frac{3}{16}\langle\hat{f}_z\rangle^2\right)\epsilon_{\bbox{k}
}
+\frac{\beta}{64}\langle\hat{f}_z\rangle^2
\right]
\right\}^\frac{1}{2}.
\label{cyclic1}
\end{eqnarray}
These excitation energies are always positive semidefinite and massive
in the presence of magnetic field.

We note that Eq.~(\ref{cyclic1}) has one gapless (but not massless)
mode in the presence of external magnetic field.
Taking the limit $|\bbox{k}|\rightarrow\bbox{0}$ of Eq.~(\ref{cyclic1}),
we obtain
\begin{eqnarray}
E_{\pm1}^{\rm C}
=\left(\frac{4}{\langle\hat{f}_z\rangle}-
\frac{\langle\hat{f}_z\rangle}{2}\right)\epsilon_{\bbox{k}}+\frac{\beta}{4}
\langle\hat{f}_z\rangle \ \ {\rm and} \ \
\left[\left(\langle\hat{f}_z\rangle-\frac{4}{\langle\hat{f}_z\rangle}
\right)^2+1\right]^\frac{1}{2}
\epsilon_{\bbox{k}}.
\label{cyclic2}
\end{eqnarray}
However, both of these modes become massless in the absence of magnetic 
field. In fact, Eq.~(\ref{cyclic1}) then reduces to
\begin{eqnarray}
E_{\pm1}^{\rm C}
&=&
\sqrt{
\epsilon_{\bbox{k}}(\epsilon_{\bbox{k}}+4c_1n)}.
\label{cyclicpm10}
\end{eqnarray}

\subsubsection{The $m=\pm2,0$ coupled modes}

The equations of motion governing the $m=\pm2,0$ coupled modes are given by
\begin{eqnarray}
i\hbar\frac{d}{dt}\hat{X}_+&=&
(A_2-(\alpha+\beta)\zeta_2^2)\hat{X}_--\gamma|\zeta_{-2}\zeta_0|\hat{Y}_-+
\gamma\zeta_0^2\hat{Z}_-/2, \nonumber \\
i\hbar\frac{d}{dt}\hat{X}_-&=&
(A_2+(\alpha+\beta)\zeta_2^2)\hat{X}_++(2\alpha\zeta_2-\gamma\zeta_{-2})
\zeta_0\hat{Y}_++(\alpha-\beta+\gamma/2)\zeta_0^2\hat{Z}_+, \nonumber \\
i\hbar\frac{d}{dt}\hat{Y}_+&=&
-\gamma|\zeta_{-2}\zeta_0|\hat{X}_-+(A_0-\alpha\zeta_0^2)\hat{Y}_--\gamma|
\zeta_2\zeta_0|\hat{Z}_-, \nonumber \\
i\hbar\frac{d}{dt}\hat{Y}_-&=&
(2\alpha\zeta_2-\gamma\zeta_{-2})\zeta_0\hat{X}_++(A_0+\alpha\zeta_0^2)
\hat{Y}_++(2\alpha\zeta_2-\gamma\zeta_2)\zeta_0\hat{Z}_+, \nonumber \\
i\hbar\frac{d}{dt}\hat{Z}_+&=&\gamma\zeta_0^2\hat{X}_-/2
-\gamma|\zeta_{-2}\zeta_0|\hat{Y}_-+(A_{-2}-(\alpha+\beta)\zeta_{-2}^2)
\hat{Z}_-, \nonumber \\
i\hbar\frac{d}{dt}\hat{Z}_-&=&
(\alpha-\beta+\gamma/2)\zeta_0^2\hat{X}_++(2\alpha\zeta_2-\gamma\zeta_2)
\zeta_0\hat{Y}_++(A_{-2}+(\alpha+\beta))\hat{Z}_+, \nonumber
\end{eqnarray}
where
$\hat{X}_\pm\equiv \hat{a}_{{\bbox{k}},2}
\pm\hat{a}_{-{\bbox{k}},2}^\dagger$,
$\hat{Y}_\pm\equiv \hat{a}_{{\bbox{k}},0}
\pm\hat{a}_{-{\bbox{k}},0}^\dagger$, and
$\hat{Z}_\pm\equiv
\hat{a}_{{\bbox{k}},-2}\pm\hat{a}_{-{\bbox{k}},-2}^\dagger$.
Substituting the expressions for $\hat{X}_-,\hat{Y}_-$ and $\hat{Z}_-$ into
those for $\hat{X}_+,\hat{Y}_+$ and $\hat{Z}_+$, we obtain the equations of
motion for the latter set, which reduces to the cubic equation and therefore
can be
solved analytically.
The result is given by
\begin{eqnarray}
E^{\rm C}_{\pm2,0}=\epsilon_{\bbox{k}}+\gamma, \ \
\left\{\epsilon_{\bbox{k}}\left[\epsilon_{\bf
k}+\alpha+\frac{4+\langle\hat{f}_z\rangle^2}{8}\beta\pm\sqrt{\alpha^2-\left(
1-\frac{3}{4}\langle\hat{f}_z\rangle^2\right)\alpha\beta+\frac{(4+\langle
\hat{f}_z\rangle^2)^2}{64}\beta^2}
\right]\right\}^\frac{1}{2},
\label{cyclicpm2}
\end{eqnarray}
where we recall that 
$\alpha\equiv c_0n$, $\beta\equiv 4c_1n$, $\gamma\equiv 2c_2n/5$.
The second solutions in Eq.~(\ref{cyclicpm2}) are always positive semidefinite.
The positivity of the first solution is guaranteed by the condition
$c_2>0$.
We note that the first solution is massive and independent of 
the applied magnetic field and
that the second solutions remain massless in the presence of external
magnetic field.
The latter is a consequence of the fact that the mean-field solution is 
degenerate with respect to at least two  continuous variables, as discussed in 
Sec.~\ref{sec:CBEC}.

In the absence of external magnetic field, the results (\ref{cyclicpm2})
reduce to
\begin{eqnarray}
E^{\rm C}_{\pm2,0}=\epsilon_{\bbox{k}}+\gamma, \ \
\sqrt{\epsilon_{\bbox{k}}(\epsilon_{\bbox{k}}+2\alpha)}, \ \
\sqrt{\epsilon_{\bbox{k}}(\epsilon_{\bbox{k}}+\beta)}.
\label{cyclicpm20}
\end{eqnarray}
It can be shown from the general analytic solutions that these results
are valid up to the first order in magnetization $\langle\hat{f}_z\rangle$.

While we have been unable to complete the analysis of the cyclic phase 
except for the case of Eq.~(\ref{cyclic_mfs2}), we would like to 
point out that the excitation spectrum of the cyclic phase always includes
the first solution in Eq.~(\ref{cyclicpm2}), 
even when the mean-field solution is not given by Eq.~(\ref{cyclic_mfs2}).
This can be seen directly by writing down the equation of motion
for $\hat{a}_{\bbox{k}m}$ and  $\hat{a}_{-\bbox{k}m}^\dagger$ and the
corresponding eigenvalue equation. It can then be seen that  
$\epsilon_{\bbox{k}}+\gamma$ is a solution to this equation.

\section{Conclusions}
\label{sec:Conclusions}

In this paper we have studied quantum spin correlations and magnetic
response of spin-2 Bose-Einstein condensates (BECs) in a mesoscopic regime,
and low-lying excitation spectra of each phase of spin-2 BECs in the
thermodynamic regime.

The ground states of spin-2 BECs have three distinct phases: ferromagnetic
(FM), antiferromagnetic (AF) and cyclic (C) phases. The former two phases
appear also in spin-1 BECs, while the last phase is unique to spin-2 BECs.
The building block of the AF phase is spin-singlet pairs and that of the
C phase is spin-singlet trios. These many-body features usually elude
mean-field treatments that are based on the order parameter derived from
the single-particle density matrix.

Bose symmetry restricts possible building blocks of spin-2 BECs.
This can be summarized in terms of the annihilation operator
$\hat{A}^{(n)}_f$ of $n$-bosons having total spin $f$.
The fundamental building block is not unique, but one minimal set is
$\hat{A}^{(1)}_2$, $\hat{A}^{(2)}_0$, $\hat{A}^{(2)}_2$, $\hat{A}^{(3)}_0$,
and $\hat{A}^{(3)}_3$. Bose statistics does not allow units such
as $\hat{A}^{(2)}_1$ and $\hat{A}^{(3)}_1$. The unit $\hat{A}^{(3)}_3$
is required to represent a state with odd values of the total spin.

We have investigated quantum spin correlations and magnetic response
in the mesoscopic regime.
Under the assumption that the system is so tightly confined that the
spatial degrees of freedom are frozen, we derived the exact many-body
ground states which are expressed in terms of the minimal set of
creation operators 
$\hat{A}^{(1)\dagger}_2$, $\hat{A}^{(2)\dagger}_0$,
$\hat{A}^{(2)\dagger}_2$,
$\hat{A}^{(3)\dagger}_0$, and $\hat{A}^{(3)\dagger}_3$.
These pairwise and trio-wise units help us understand the complicated
response of the magnetization to the applied magnetic field, which
stems from the fact that several values of the magnetization
cannot be constructed from such units and are hence forbidden.
In addition to the quantization of the magnetization to discrete
values,
several new features which elude mean-field treatments are found,
such as a sudden jump from the minimum to the maximum magnetization,
and robustness of the minimum-magnetization state against a small
increase in the applied magnetic field until it starts to show a linear
response. The average Zeeman level populations for the AF-phase
ground states were calculated, showing that $m=0,\pm 1$ populations,
which stay zero in MFT, vary sensitively to the applied
magnetic field.

We have also examined low-lying excitation spectra using the Bogoliubov
approximation. The excitation spectra of FM and AF phases are similar
to those of the spin-1 case~\cite{OM,Ueda}.
In the FM phase, the spectrum
consists of one massless mode (\ref{FBog}) reflecting the global
gauge invariance and four single-particle modes (\ref{FS1})-(\ref{FSM2})
whose energy gaps are generated by the Zeeman shifts as well as
mean-field interactions.
In the AF phase, the spectrum includes two massless modes (\ref{dis2})
due to the global gauge invariance and the rotational symmetry about the
spin quantization axis. The remaining three are also Bogoliubov modes, but
they all become massive in the presence of magnetic field due to the
Zeeman shifts.
In the C phase, the spectrum have at least two  massless modes (the second
term in Eq.~(\ref{cyclicpm2})) by the same reasons as in the AF phase.
The spectrum includes one peculiar single-particle mode (the first
term in Eq.~(\ref{cyclicpm2})) whose energy gap depends solely on the
spin-dependent interactions and is insensitive to the
applied magnetic field.
In addition, the spectrum has one gapless mode
(the second term in Eq.~(\ref{cyclic2})) whose mass depends only on
magnetization $\langle\hat{f}_z\rangle$ and vanishes at zero magnetic field.
The remaining mode (the first term in Eq.~(\ref{cyclic2})) is a
Bogoliubov mode which becomes massive in the presence of the magnetic field.

In the present paper we have studied only static properties of spin-2 BEC.
With the very rich phenomena that we have found here, we may very well expect
that much more remains to be revealed in their dynamics.

\acknowledgments

This work was supported by a Grant-in-Aid for Scientific Research
(Grant No. 11216204) by the Ministry of Education, Science, Sports,
and Culture of Japan, and by the Toray Science Foundation.
M.U. acknowledges the hospitality of the Aspen Center for Physics,
where part of this work was carried out.

\appendix

\section{The order parameter of spin-2 BEC}
\label{app:Characterization}

The order parameter ${\bf \zeta}$ of a spin-2 BEC has the same structure as
that of the d-wave superconductor which was examined by
Mermin~\cite{Mermin}.
We here recapitulate as much of it as is relevant to the our theory.
The spin part of the order parameter,
$\Psi$, of a spin-2 BEC is described as
a function of the azimuthal angle $\theta$ and the radial angle $\phi$, and
may be expanded in terms of the spherical harmonics of
rank 2 $Y_{2}^m(\theta,\phi)$ as
\begin{eqnarray}
\Psi=\sum_{m=-2}^2\zeta_mY_{2}^m,
\label{chi1}
\end{eqnarray}
where $\zeta_m$ obeys the normalization condition (\ref{norm}).
The angle dependence of $Y_{2}^m$ may be expressed in terms of
components of a three-dimensional unit vector:
$\hat{\bf n}^T=(k_x,k_y,k_z)\equiv
(\sin\theta\cos\phi,\sin\theta\sin\phi,\cos\theta)$
as follows
\begin{eqnarray}
Y_{2}^{\pm2}=\sqrt{\frac{15}{32\pi}}(k_x\pm ik_y)^2, \ \
Y_{2}^{\pm1}=\mp\sqrt{\frac{15}{8\pi}}(k_x\pm ik_y)k_z, \ \
Y_{2}^{0}   =\sqrt{\frac{5}{16\pi}}(2k_z^2-k_x^2-k_y^2).
\end{eqnarray}
Substituting these into Eq.~(\ref{chi1}), we obtain
\begin{eqnarray}
\Psi=\sqrt{\frac{15}{8\pi}}\hat{\bf n}^T{\bf M}\hat{\bf n},
\label{chi2}
\end{eqnarray}
where
\begin{eqnarray}
{\bf M}=\frac{1}{2}
\left(
\begin{array}{ccc}
\zeta_2+\zeta_{-2}-\sqrt{\frac{2}{3}}\zeta_0 & i(\zeta_2-\zeta_{-2})
& -\zeta_1+\zeta_{-1}     \\
i(\zeta_2-\zeta_{-2})  & -\zeta_2-\zeta_{-2}-\sqrt{\frac{2}{3}}\zeta_0
& -i(\zeta_1+\zeta_{-1})  \\
-\zeta_1+\zeta_{-1} & -i(\zeta_1+\zeta_{-1}) & 2\sqrt{\frac{2}{3}}\zeta_0
\end{array}
\right).
\label{M}
\end{eqnarray}
The order parameter is thus characterized by a $3\times3$ traceless matrix
Tr${\bf M}=0$ with unit normalization
\begin{eqnarray}
{\rm Tr}({\bf M}^*{\bf M})=1.
\label{MM}
\end{eqnarray}
We may exploit the freedom of the gauge invariance to choose the 
overall phase
so that the real part of Tr${\bf M}^2$ vanishes.
\begin{eqnarray}
{\rm Re}{\rm Tr}{\bf M}^2=0.
\label{Re}
\end{eqnarray}
Let the real and imaginary parts of ${\bf M}$ be ${\bf X}$ and ${\bf Y}$,
respectively. It follows from Eqs.~(\ref{MM}) and (\ref{Re}) that
\begin{eqnarray}
{\rm Tr}{\bf X}^2={\rm Tr}{\bf Y}^2=\frac{1}{2}.
\label{XX}
\end{eqnarray}
Because ${\bf B}$ is traceless, so can be ${\bf X}$ and ${\rm Y}$ traceless.
${\bf X}$ and ${\rm Y}$ do not commute, so they cannot be diagonalized
simultaneously. We follow Mermin to take a representation in which ${\bf X}$
is diagonal. Then the diagonal elements of ${\bf X}$ become
\begin{eqnarray}
x_n=\sqrt{\frac{1}{3}}\sin\left(\theta+\frac{2\pi}{3}n\right).
\label{XD}
\end{eqnarray}
The matrix elements of ${\bf Y}$ are given by
\begin{eqnarray}
& &
Y_{nn}=\sqrt{\frac{1}{3}}\sin\left(\phi+\frac{2\pi}{3}n\right)\sin\chi,
\ \
Y_{23}=Y_{32}=-\frac{1}{2}\sin\delta\cos\psi\cos\chi, \nonumber \\
& & Y_{31}=Y_{13}=-\frac{1}{2}\sin\delta\sin\psi\cos\chi, \ \
Y_{12}=Y_{21}= \frac{1}{2}\cos\delta\cos\chi.
\label{YY}
\end{eqnarray}
Substituting Eqs.~(\ref{XX}) and (\ref{YY}) into ${\bf M}={\bf X}+i{\bf Y}$
and comparing it with Eq.~(\ref{M}), we obtain the following
characterization
of the order parameter.
\begin{eqnarray}
& & \zeta_{\pm2}=\frac{1}{2}
(\cos\theta\pm\cos\delta\cos\chi+i\cos\phi\sin\chi)
\nonumber \\
& & \zeta_{\pm1}=\frac{1}{2}\cos\chi\sin\delta e^{\pm i\psi}
\nonumber \\
& & \zeta_0=\frac{1}{\sqrt{2}}(\sin\theta+i\sin\chi\sin\phi).
\label{order}
\end{eqnarray}

\section{Calculation of Zeeman-level populations}
\label{app:zeeman}

\subsection{Derivation of Eq.~(\protect\ref{formula_zeeman1})}

Let us first write $\langle\hat{a}_m^\dagger\hat{a}_m\rangle$ as
\begin{equation}
\langle\hat{a}_m^\dagger\hat{a}_m\rangle
=\frac{\langle \phi_m|(\hat{\cal S}_-)^{N_{\rm S}}
(\hat{\cal S}_+)^{N_{\rm S}} |\phi_m\rangle}
{\langle \phi|(\hat{\cal S}_-)^{N_{\rm S}}
(\hat{\cal S}_+)^{N_{\rm S}} |\phi\rangle}
-1
\label{am_by_norms}
\end{equation}
with $|\phi_m\rangle\equiv\hat{a}_m^\dagger|\phi\rangle$.
What we need is thus a general formula for calculating
$\langle \psi|(\hat{\cal S}_-)^l
(\hat{\cal S}_+)^l |\psi\rangle$.
Let us first consider the decomposition of $|\psi\rangle$ into a sum of
eigenstates for $\hat{\cal S}^2$, such that
\begin{equation}
|\psi\rangle=\sum_{k=0}^{\lfloor n/2 \rfloor}
(\hat{\cal S}_+)^k|\psi_k\rangle,
\label{decomp_psi_S}
\end{equation}
where $|\psi_k\rangle$ is an unnormalized simultaneous eigenstate of
$\{\hat{\cal S}^2,
\hat{\cal S}_z\}$ with eigenvalues $S(S-1)$ and $S=[2(n-2k)+5]/4$,
respectively.
It follows from Eqs.~(\ref{eigen_S}) and (\ref{def_N2})
that
$N_s=0$ and $N_0=n-2k$. Here $n$ is the number of bosons in
$|\psi\rangle$, and $\lfloor x \rfloor$ denotes the largest integer that
is not  larger than $x$.
Let us define $\omega_j$ and $\mu_j$ such that
\begin{equation}
\omega_j\equiv \frac{\langle \psi|(\hat{\cal S}_+)^j
(\hat{\cal S}_-)^j|\psi\rangle}
{\langle \psi|\psi\rangle}
\label{def_omega}
\end{equation}
and
\begin{equation}
\mu_j\equiv \frac{\langle \psi_j|(\hat{\cal S}_-)^j
(\hat{\cal S}_+)^j|\psi_j\rangle}
{\langle \psi|\psi\rangle}
=(n-2j;j)
\frac{\langle \psi_j|\psi_j\rangle}
{\langle \psi|\psi\rangle},
\end{equation}
where we have used Eq.~(\ref{formula_S_plus}) and
defined the coefficient $(a;b)$ as
\begin{equation}
(a;b)\equiv \frac{b!(b+a+3/2)!}{(a+3/2)!}
\label{coeff_s_plus}.
\end{equation}
By definition, $\omega_0=1$ and $\sum_j \mu_j=1$.
Substituting Eq.~(\ref{def_omega}) into
Eq.~(\ref{decomp_psi_S}) yields
\begin{equation}
\omega_j=(n-2j;j)\mu_j+\sum_{k=j+1}^{\lfloor n/2
\rfloor}\mu_k\frac{(n-2k;k)}{(n-2k;k-j)},
\end{equation}
or equivalently,
\begin{equation}
\mu_j=\frac{\omega_j}{(n-2j;j)}-\sum_{k=j+1}^{\lfloor n/2
\rfloor}\mu_k\frac{(n-2k;k)}{(n-2j;j)(n-2k;k-j)}.
\label{mu_by_omega}
\end{equation}
Using this relation recursively, we can calculate
$\mu_j$ as a function of $\{\omega_{k}\} (k=j,j+1,\ldots,\lfloor n/2
\rfloor)$.
On the other hand, multiplying $(\hat{\cal S}_+)^l$ on both sides of
Eq.~(\ref{decomp_psi_S}) and taking norms, we obtain
\begin{equation}
\frac{\langle \psi|(\hat{\cal S}_-)^l
(\hat{\cal S}_+)^l |\psi\rangle}
{\langle \psi|\psi\rangle}
=\sum_{k=0}^{\lfloor n/2 \rfloor}\frac{(n-2k;l+k)}{(n-2k;k)}\mu_k.
\label{norm_by_mu}
\end{equation}

Using this formula, we can evaluate Eq.~(\ref{am_by_norms}).
   If we apply $|\psi\rangle=|\phi\rangle$, $n=s$, and $l=N_{\rm S}$ to the
relations (\ref{def_omega}), (\ref{mu_by_omega}), and
(\ref{norm_by_mu}), and noting that $\hat{\cal S}_-|\phi\rangle=0$,
   we have
$\omega_0=\mu_0=1$, $\omega_j=\mu_j=0 (j>0)$,
and
\begin{equation}
\frac{\langle \phi|(\hat{\cal S}_-)^{N_{\rm S}}
(\hat{\cal S}_+)^{N_{\rm S}} |\phi\rangle}
{\langle \phi|\phi\rangle}
=(s;{N_{\rm S}}).
\label{norm_phi}
\end{equation}

For $|\phi\rangle_m$,
we have
$
\hat{\cal S}_-|\phi_m\rangle=
[\hat{\cal S}_-,\hat{a}_m^\dagger]|\phi\rangle
=(-1)^m\hat{a}_{-m}|\phi\rangle
$
and $(\hat{\cal S}_-)^2|\phi_m\rangle=0$.
Using these when we apply $|\psi\rangle=|\phi_m\rangle$, $n=s+1$, and
$l=N_{\rm S}$ to the relations (\ref{def_omega}),
(\ref{mu_by_omega}), and (\ref{norm_by_mu}), we have
$\omega_1=\langle \phi|\hat{a}_{-m}^\dagger\hat{a}_{-m}
|\phi\rangle/\langle \phi_m|\phi_m\rangle$,
$\omega_j=\mu_j=0 (j>1)$,
$\mu_1=\omega_1/(s-1;1)$,
$\mu_0=1-\mu_1$, and
\begin{equation}
\frac{\langle \phi_m|(\hat{\cal S}_-)^{N_{\rm S}}
(\hat{\cal S}_+)^{N_{\rm S}} |\phi_m\rangle}
{\langle \phi_m|\phi_m\rangle}
=(s+1;{N_{\rm S}}) \mu_0+
\frac{(s-1;{N_{\rm S}}+1)}{(s-1;1)}\mu_1.
\label{norm_phi_prime}
\end{equation}
Combining Eqs.~(\ref{am_by_norms}), (\ref{norm_phi}), and
(\ref{norm_phi_prime}), we obtain
\begin{equation}
\langle\hat{a}_m\hat{a}_m^\dagger\rangle
=\frac{(s+1;{N_{\rm S}})} {(s;{N_{\rm S}})}
\left(\langle\hat{a}_m\hat{a}_m^\dagger\rangle_0-\frac
{\langle\hat{a}_{-m}^\dagger\hat{a}_{-m}\rangle_0}
{(s-1;1)}\right)
+\frac{(s-1;{N_{\rm S}}+1)}{(s;{N_{\rm S}})(s-1;1)^2}
\langle\hat{a}_{-m}^\dagger\hat{a}_{-m}\rangle_0,
\end{equation}
where $\langle\cdots\rangle_0\equiv
\langle \phi|\cdots|\phi\rangle/\langle \phi|\phi\rangle$.
Substituting Eq.(\ref{coeff_s_plus}), we obtain
Eq.~(\ref{formula_zeeman1}).

\subsection{Exact forms of Zeeman populations}

Here we give the results of Zeeman populations
$\langle\hat{a}_m^\dagger\hat{a}_m\rangle$ for
the states $(\hat{A}^{(2)\dagger}_0)^{N_{\rm S}}
(\hat{a}^\dagger_2)^{n_{12}}
(\hat{A}^{(2)\dagger}_2)^{n_{22}}
(\hat{A}^{(3)\dagger}_3)^{n_{33}}|{\rm vac}\rangle$
with $n_{22}=0,1$ and $n_{33}=0,1$.

i) $n_{22}=0$ and $n_{33}=0$. We take
$|\phi\rangle=(\hat{a}^\dagger_2)^{n_{12}}|{\rm vac}\rangle$
and $s=n_{12}$. Then,
$\langle\hat{a}_2^\dagger\hat{a}_2\rangle_0=n_{12}$
and $\langle\hat{a}_m^\dagger\hat{a}_m\rangle_0=0$ for $m<2$.
Putting these into the formula (\ref{formula_zeeman1}),
we obtain
\begin{equation}
\langle\hat{a}_2^\dagger\hat{a}_2\rangle
=n_{12}+N_{\rm S}-
\frac{3N_{\rm S}}{2n_{12}+5},
\end{equation}
\begin{equation}
\langle\hat{a}_1^\dagger\hat{a}_1\rangle
=\langle\hat{a}_0^\dagger\hat{a}_0\rangle
=\langle\hat{a}_{-1}^\dagger\hat{a}_{-1}\rangle
=
\frac{2N_{\rm S}}{2n_{12}+5},
\end{equation}
\begin{equation}
\langle\hat{a}_{-2}^\dagger\hat{a}_{-2}\rangle
=N_{\rm S}-
\frac{3N_{\rm S}}{2n_{12}+5}.
\end{equation}

ii) $n_{22}=1$ and $n_{33}=0$. We take
$|\phi\rangle=(\hat{a}^\dagger_2)^{n_{12}}
[\sqrt{3}(\hat{a}_1)^2-2\sqrt{2}\hat{a}_2\hat{a}_0]^\dagger|{\rm
vac}\rangle$
and $s=n_{12}+2$. Then,
\begin{equation}
\langle\hat{a}_2^\dagger\hat{a}_2\rangle_0
=n_{12}+1-
\frac{3}{4n_{12}+7},
\end{equation}
\begin{equation}
\langle\hat{a}_1^\dagger\hat{a}_1\rangle_0
=
\frac{6}{4n_{12}+7},
\end{equation}
\begin{equation}
\langle\hat{a}_0^\dagger\hat{a}_0\rangle_0
=1-
\frac{3}{4n_{12}+7},
\end{equation}
and $\langle\hat{a}_m^\dagger\hat{a}_m\rangle_0=0$ for $m<0$.
Putting these into the formula (\ref{formula_zeeman1}),
we obtain
\begin{equation}
\langle\hat{a}_2^\dagger\hat{a}_2\rangle
=n_{12}+N_{\rm S}-
\frac{5N_{\rm S}}{2n_{12}+9}
-\left(\frac{2N_{\rm S}}{2n_{12}+9}+1\right)\frac{3}{4n_{12}+7}+1,
\end{equation}
\begin{equation}
\langle\hat{a}_1^\dagger\hat{a}_1\rangle
=
\frac{2N_{\rm S}}{2n_{12}+9}+
\left(\frac{2N_{\rm S}}{2n_{12}+9}+1\right)
\frac{6}{4n_{12}+7},
\end{equation}
\begin{equation}
\langle\hat{a}_0^\dagger\hat{a}_0\rangle
=
\frac{6N_{\rm S}}{2n_{12}+9}-
\left(\frac{4N_{\rm S}}{2n_{12}+9}+1\right)
\frac{3}{4n_{12}+7}+1,
\end{equation}
\begin{equation}
\langle\hat{a}_{-1}^\dagger\hat{a}_{-1}\rangle
=
\frac{2N_{\rm S}}{2n_{12}+9}
\left(1+\frac{6}{4n_{12}+7}\right),
\end{equation}
\begin{equation}
\langle\hat{a}_{-2}^\dagger\hat{a}_{-2}\rangle
=N_{\rm S}-
\frac{N_{\rm S}}{2n_{12}+9}
\left(5+\frac{6}{4n_{12}+7}\right).
\end{equation}

iii) $n_{22}=0$ and $n_{33}=1$. We take
$|\phi\rangle=(\hat{a}^\dagger_2)^{n_{12}}
[(\hat{a}_1)^3-\sqrt{6}\hat{a}_2\hat{a}_1\hat{a}_0
+2(\hat{a}_2)^2\hat{a}_{-1}]^\dagger|{\rm vac}\rangle$
and $s=n_{12}+3$. Then,
\begin{equation}
\langle\hat{a}_2^\dagger\hat{a}_2\rangle_0
=n_{12}+2-
\frac{3(n_{12}+3)}{(n_{12}+2)(2n_{12}+5)},
\end{equation}
\begin{equation}
\langle\hat{a}_1^\dagger\hat{a}_1\rangle_0
=
\frac{3(n_{12}+4)}{(n_{12}+2)(2n_{12}+5)},
\end{equation}
\begin{equation}
\langle\hat{a}_0^\dagger\hat{a}_0\rangle_0
=\frac{3(n_{12}+1)}{(n_{12}+2)(2n_{12}+5)},
\end{equation}
\begin{equation}
\langle\hat{a}_{-1}^\dagger\hat{a}_{-1}\rangle_0
=1-\frac{3}{2n_{12}+5},
\end{equation}
and
$\langle\hat{a}_{-2}^\dagger\hat{a}_{-2}\rangle_0=0$.
Putting these into the formula (\ref{formula_zeeman1}),
we obtain
\begin{equation}
\langle\hat{a}_2^\dagger\hat{a}_2\rangle
=n_{12}+N_{\rm S}-
\frac{5N_{\rm S}}{2n_{12}+11}-
\left(\frac{2N_{\rm S}}{2n_{12}+11}+1\right)
\frac{3(n_{12}+3)}{(n_{12}+2)(2n_{12}+5)}+2,
\end{equation}
\begin{equation}
\langle\hat{a}_1^\dagger\hat{a}_1\rangle
=
\frac{4N_{\rm S}}{2n_{12}+11}+
\left(\frac{12N_{\rm S}}{2n_{12}+11}+3(n_{12}+4)\right)
\frac{1}{(n_{12}+2)(2n_{12}+5)},
\end{equation}
\begin{equation}
\langle\hat{a}_0^\dagger\hat{a}_0\rangle
=\frac{2N_{\rm S}}{2n_{12}+11}+
\left(\frac{4N_{\rm S}}{2n_{12}+11}+1\right)
\frac{3(n_{12}+1)}{(n_{12}+2)(2n_{12}+5)},
\end{equation}
\begin{equation}
\langle\hat{a}_{-1}^\dagger\hat{a}_{-1}\rangle
=
\frac{4N_{\rm S}}{2n_{12}+11}+
\left(\frac{12N_{\rm S}}{2n_{12}+11}-3(n_{12}+2)\right)
\frac{1}{(n_{12}+2)(2n_{12}+5)}+1,
\end{equation}
\begin{equation}
\langle\hat{a}_{-2}^\dagger\hat{a}_{-2}\rangle
=N_{\rm S}-
\frac{N_{\rm S}}{2n_{12}+11}
\left(5+\frac{6(n_{12}+3)}{(n_{12}+2)(2n_{12}+5)}\right)
.
\end{equation}

iv) $n_{22}=1$ and $n_{33}=1$. We take
$|\phi\rangle=(\hat{a}^\dagger_2)^{n_{12}}
[\sqrt{3}(\hat{a}_1)^5-5\sqrt{2}\hat{a}_2(\hat{a}_1)^3\hat{a}_0
+4\sqrt{3}(\hat{a}_2)^2\hat{a}_1(\hat{a}_0)^2
+2\sqrt{3}(\hat{a}_2)^2(\hat{a}_1)^2\hat{a}_{-1}
-4\sqrt{2}(\hat{a}_2)^3\hat{a}_0\hat{a}_{-1}]^\dagger|{\rm vac}\rangle$
and $s=n_{12}+5$. Then,
\begin{equation}
\langle\hat{a}_2^\dagger\hat{a}_2\rangle_0
=n_{12}+3-
\frac{30(n_{12}+4)^2}{(n_{12}+3)(2n_{12}+7)(4n+13)},
\end{equation}
\begin{equation}
\langle\hat{a}_1^\dagger\hat{a}_1\rangle_0
=
\frac{9(4n_{12}^2+37n_{12}+83)}{(n_{12}+3)(2n_{12}+7)(4n+13)},
\end{equation}
\begin{equation}
\langle\hat{a}_0^\dagger\hat{a}_0\rangle_0
=1+\frac{18(n_{12}^2+3n_{12}-3)}{(n_{12}+3)(2n_{12}+7)(4n+13)},
\end{equation}
\begin{equation}
\langle\hat{a}_{-1}^\dagger\hat{a}_{-1}\rangle_0
=1-\frac{3(8n_{12}^2+49n_{12}+71)}{(n_{12}+3)(2n_{12}+7)(4n+13)},
\end{equation}
and
$\langle\hat{a}_{-2}^\dagger\hat{a}_{-2}\rangle_0=0$.
Putting these into the formula (\ref{formula_zeeman1}),
we obtain
\begin{equation}
\langle\hat{a}_2^\dagger\hat{a}_2\rangle
=n_{12}+N_{\rm S}-
\frac{7N_{\rm S}}{2n_{12}+15}-
\left(\frac{2N_{\rm S}}{2n_{12}+15}+1\right)
\frac{30(n_{12}+4)^2}{(n_{12}+3)(2n_{12}+7)(4n+13)}+3,
\end{equation}
\begin{equation}
\langle\hat{a}_1^\dagger\hat{a}_1\rangle
=
\frac{4N_{\rm S}}{2n_{12}+15}+
\left(4N_{\rm S}(n_{12}+8)-\frac{124N_{\rm
S}}{2n_{12}+15}+3(4n_{12}^2+37n_{12}+83)\right)
\frac{3}{(n_{12}+3)(2n_{12}+7)(4n+13)},
\end{equation}
\begin{equation}
\langle\hat{a}_0^\dagger\hat{a}_0\rangle
=\frac{6N_{\rm S}}{2n_{12}+15}+
\left(\frac{4N_{\rm S}}{2n_{12}+15}+1\right)
\frac{18(n_{12}^2+3n_{12}-3)}{(n_{12}+3)(2n_{12}+7)(4n+13)}+1,
\end{equation}
\begin{equation}
\langle\hat{a}_{-1}^\dagger\hat{a}_{-1}\rangle
=
\frac{4N_{\rm S}}{2n_{12}+15}+
\left(4N_{\rm S}(n_{12}+8)-\frac{124N_{\rm
S}}{2n_{12}+15}-(8n_{12}^2+49n_{12}+71)\right)
\frac{3}{(n_{12}+3)(2n_{12}+7)(4n+13)}+1,
\end{equation}
\begin{equation}
\langle\hat{a}_{-2}^\dagger\hat{a}_{-2}\rangle
=N_{\rm S}-
\frac{N_{\rm S}}{2n_{12}+15}
\left(7+\frac{60(n_{12}+4)^2}{(n_{12}+3)(2n_{12}+7)(4n+13)}\right)
.
\end{equation}

\end{document}